\documentclass[fleqn,usenatbib]{mnras}

\usepackage{newtxtext,newtxmath}

\usepackage[T1]{fontenc}

\DeclareRobustCommand{\VAN}[3]{#2}
\let\VANthebibliography\thebibliography
\def\thebibliography{\DeclareRobustCommand{\VAN}[3]{##3}\VANthebibliography}

\usepackage{graphicx}	
\usepackage{amsmath}	
\usepackage[normalem]{ulem}

\usepackage{color}

\def \fexii {Fe\,{\sc xii}}
\def \fexiv {Fe\,{\sc xiv}}
\def \fexv {Fe\,{\sc xv}}

\title[Non-thermal broadening]{Non-thermal broadening of coronal lines in a 3D MHD loop model}

\author[C. A. Breu et al.]{
C.A. Breu,$^{1,2}$\thanks{E-mail: cab42@st-andrews.ac.uk (University of St. Andrews)}, H. Peter,$^{2}$ S.K. Solanki,$^{2,3}$ R. Cameron,$^{2}$ I. De Moortel,$^{1,4}$ 
\\
$^{1}$School of Mathematics and Statistics, University of St Andrews, St Andrews, Fife KY16 9SS, UK\\
$^{2}$Max Planck Institute for Solar System Research,  Justus-von-Liebig-Weg 3, 37077 G\"ottingen, Germany\\
$^{3}$School of Space Research, Kyung Hee University, Yongin, Gyeonggi 446-701, Republic of Korea\\
$^{4}$Rosseland Centre for Solar Physics, University of Oslo, PO Box 1029 Blindern, NO-0315 Oslo, Norway
}

\date{Accepted XXX. Received YYY; in original form ZZZ}

\pubyear{2024}

\begin{document}
\label{firstpage}
\pagerange{\pageref{firstpage}--\pageref{lastpage}}
\maketitle

\begin{abstract}
Observed spectral profiles of emission lines from the corona are found to have widths exceeding the thermal line width.
To investigate the physical mechanism, we run a 3D MHD model of a single, straightened loop in which we partially resolve turbulent motions that form in response to the driving by self-consistently evolving magneto-convection in the photosphere.
The convective motions shear and twist the magnetic field lines, leading to heating. From the model we synthesize spectral profiles of emission lines forming at temperatures around and above 1 MK. The coronal heating process generates a range of velocity amplitudes and directions structured on a scale much smaller than the resolving power of current instruments, leading to a broadening of the spectral lines.
 Our model includes the mass exchange between corona and chromosphere,  thus we also capture flows parallel to the loop axis.
We find that the spectral lines show a non-thermal line broadening roughly consistent with observations for a viewing angle perpendicular to the axis. The broadening through field-parallel flows is comparable, although slightly smaller. The line broadening is independent of the instrument resolution for a perpendicular line-of-sight.  We can connect the non-thermal line broadening to heating events and flows. While small-scale velocities along the line-of-sight are mainly responsible for the broadening observed perpendicular to the loop, chromospheric evaporation is important for the line broadening observed along the loop.
The model reproduces observed values for non-thermal line widths. In the model these result from continuous driving by magnetoconvection,  without imposing driving motions or starting from an already braided field. 
\end{abstract}


\maketitle

\section{Introduction}
\label{section:intro}

Different processes can cause the broadening of emission lines.
In thermodynamic equilibrium, the particles in a plasma move according to a Maxwellian velocity distribution with a width determined by the temperature. The motion of the emitting particles results in a broadening of emission lines, the thermal broadening.
In observations of the solar corona, emission lines in the extreme ultraviolet show broadening that exceeds the thermal line width. Unresolved motions within a resolution element of the observing instrument and along the line of sight lead to non-thermal broadening of spectral lines.\\ 
The amount of this non-thermal line broadening depends on the observed solar region. Different values were found for the quiet Sun, active regions or coronal holes. It is largest in the quiet Sun with values up to $\rm{30\; km\; s^{-1}}$ \citep{1998ApJ...505..957C}. Typical values for the observed non-thermal broadening of emission lines in active regions from plasma above 1 MK are in the range of 15-20 $\mathrm{km\; s^{-1}}$ \citep{1999ApJ...513..969H,2016ApJ...820...63B}. A correlation of intensity and non-thermal line broadening was found for lines emitted in the low transition region for temperatures in the range up to $10^{5}\; \rm{K}$ \citep{1984ApJ...281..870D}. For hotter plasma in active regions or coronal loops, there is only a weak correlation between the intensity and the non-thermal line broadening \citep{1998ApJ...505..957C}.\\
The observed non-thermal broadening is independent of the instrument resolution \citep{2015ApJ...799L..12D,2016ApJ...827...99T}.\\
Potential processes to explain these unresolved motions are turbulence, quasi-periodic upflows, nanoflares, shocks or waves \citep{2015ApJ...799L..12D, 2020A&A...639A..21P}. It follows from the independence of non-thermal line broadening from the spatial resolution of the observing instrument that the process responsible for line broadening must operate along the line of sight or on scales below the highest currently available instrument resolution of a coronal or transition region spectrometer (ca. $2^{\prime\prime}$ for Hinode/EIS \citep{2007SoPh..243...19C} and $0.^{\prime\prime}33-0.^{\prime\prime}4$ for IRIS \citep{2014SoPh..289.2733D}. Otherwise one would expect that the non-thermal broadening increases with decreasing spatial instrument resolution.\\
Due to the frozen-in condition, the movement of plasma in the corona is strongly impeded in the direction perpendicular to the magnetic field. Therefore, non-thermal velocities along and across the magnetic field are likely to arise from different processes. The correlation between line width and intensity has been linked to shocks for a LOS parallel to the magnetic field, while for the perpendicular direction, small-scale twist could explain the correlation \citep{2015ApJ...799L..12D}.\\
The shapes of the spectral line profiles can provide information about heating mechanisms and mass flows in the corona. 
If a structure is observed at the limb, motions perpendicular to the magnetic field are expected to produce the dominant contribution to the non-thermal line broadening near the limb, while loops seen edge-on would have a contribution from field-aligned motions near the apex. Line widths measured in on-disk observations mainly contain components from motions along the magnetic field near the loop footpoints and motions perpendicular to the loop axis near the apex. 
The center-to-limb variation of the observed non-thermal broadening can therefore help to disentangle contributions from flows along and perpendicular to the magnetic guide field.\\
Observed line profiles of transition region lines in the quiet Sun have several components consisting of a narrow core and a broad second component that could be related to flows in the footpoint regions of large loops \citep{2000A&A...360..761P}. While \citet{1998ApJ...505..957C} did not find a significant center-to-limb variation of the non-thermal broadening for lines from plasma at temperatures between $10^{4}\; \rm{K}$ and $2\times 10^{6}\; \rm{K}$, \citet{2008ApJ...678L..67H} found that non-thermal velocities observed at the footpoints of an active region in the \fexiv\ and \fexv\ lines decrease towards the limb. They interpret deviations from a Gaussian profile in the blue line wing as strong unresolved upflows and conclude that the enhanced line broadening at loop footpoints in the disk observations is due to field-aligned flows. \citet{2019A&A...626A..98L} interpreted the blue wing of spectra measured at the footpoint of a cool, low-lying loop as plasma injection into the loop. \citet{2010A&A...521A..51P} also found blueshifted components in the profiles of the \fexv\ line at 284 \AA\ in the footpoint regions of an active region. \citet{1998A&A...337..287E} observed a broadening of the line profiles of chromospheric and transition region lines towards the limb and interpreted this as signatures of Alfv\'{e}n waves.\\
Non-thermal line broadening can arise both from the heating process itself, for example from unresolved wave motions, and from the response of the plasma to the heating. The line width is often taken as a measure for the root-mean-square velocity and thus the energy flux carried into the corona by various types of waves \citep{2012ApJ...761..138M, 2019ApJ...881...95P}. Under this assumption the non-thermal line broadening arises from the mechanism that carries energy into the corona and thus from the cause of the heating.\\
Not only wave motions may lead to increased line broadening, nanoflares producing coronal heating can also cause non-thermal broadening.
Reconnection of magnetic field lines in the atmosphere can drive bidirectional reconnection jets that lead to spectral line profiles with separate components in the wings  \citep{1993SoPh..144..217D,1997Natur.386..811I,2021NatAs...5...54A}.\\
Energy release in the corona due to reconnection could also lead to non-thermal broadening by  chromospheric evaporation \citep{2010A&A...521A..51P}, as was suggested by \citet{2006ApJ...647.1452P}. \citet{2010A&A...521A..51P}, however, found that the velocities at the footpoints are too low compared to the model by \citet{2006ApJ...647.1452P}. Furthermore, the line broadening higher up in the loop was stronger than at the loop footpoints. This scenario would be more compatible with transverse Alfv\'{e}n waves. In a model driven self-consistently by magnetoconvection, it is not straightforward to separate clearly braiding and wave heating. The simulation contains plasma flows and changes in the magnetic field on a range of timescales, therefore it is expected that motions contribute which are in both the slower (braiding) and faster (waves) regimes.\\\\
The non-thermal line broadening is underestimated in numerical simulations of active regions compared to observations for transition region lines \citep{2006ApJ...638.1086P}, although the values are closer to observations for lines formed at coronal temperatures \citep{2006ApJ...638.1086P, 2015ApJ...802....5O}. A possible reason is that the resolution in those models is not high enough. The vertical resolution in \citet{2006ApJ...638.1086P} goes down to 150 km, while the resolution in \citet{2015ApJ...802....5O} is 47.6 km in the horizontal direction and ranges from 18 km to 80 km in the vertical direction. With a higher resolution, small-scale flows with potentially higher velocities could be resolved \citep{2006ApJ...638.1086P,2020A&A...639A..21P}.\\
Several aspects of the observed characteristics of non-thermal line broadening have been successfully reproduced in a simulation of the turbulent relaxation of a magnetic braid by \citet{2020A&A...639A..21P}, including the typical observed values, independence of the spatial resolution of the observing instrument and excess emission in the line wings, which is interpreted as a signature of turbulent motion. However, in that model the simulation was started from a pre-braided magnetic field that is relaxing instead of incorporating the driving of the magnetic field by convection.
In this paper, we investigate whether observed non-thermal line widths are reproduced in a Cartesian model of an isolated flux tube including part of the convection zone layer.
\\
This paper is organized as follows. First the simulation setup and the calculation of synthetic spectra is described in Sect. \ref{section:meth}, then the results for the non-thermal line width are presented for different spatial effective resolutions of the observing instrument in Sect. \ref{section:res_ntlw} and discussed with respect to coronal heating and flows triggered by heating events in Sect. \ref{section:disc_ntlw}. We present conclusions in Sect. \ref{section:conc}.
\section{Methods}
\label{section:meth}

\subsection{The loop model}

\begin{figure}
\resizebox{\hsize}{!}{\includegraphics{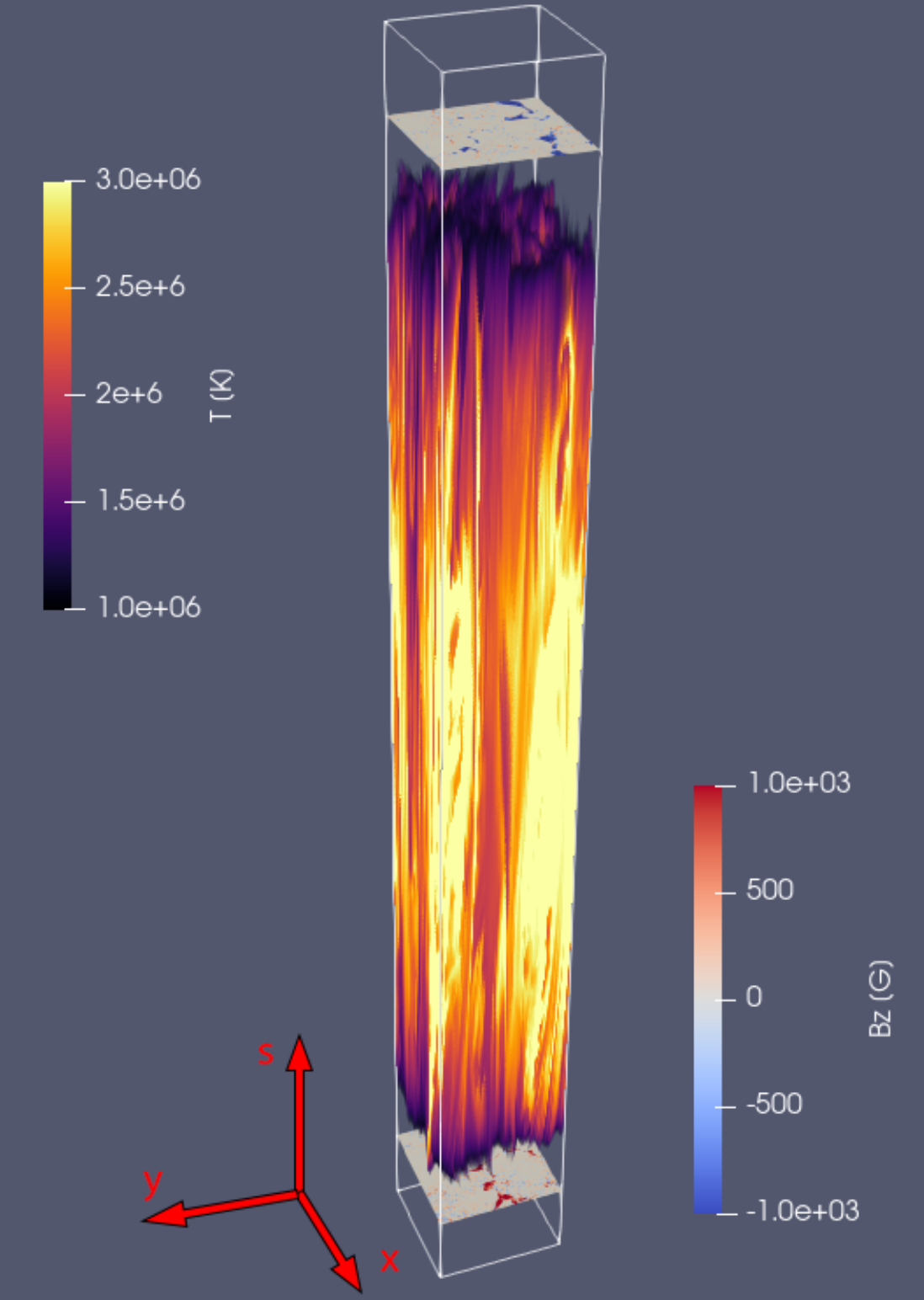}}
  \caption{Simulation setup. Volume rendering of the loop temperature in Kelvin. The slices in the photosphere at both ends of the loop show the axial component of the magnetic field. The dimension of the computational domain is 6 Mm by 6 Mm by 57 Mm with an effective loop length of 50 Mm. See Sect. \ref{section:meth}.}
  \label{fig:loop_model}
\end{figure}

We solve the compressible, resistive MHD equations with the MURaM code \citep{2003PhDT........61V, 2005A&A...429..335V} including the coronal extension \citep{2017ApJ...834...10R}. 
The coronal loop is modelled as a straightened-out magnetic flux tube with a coronal part spanning the space between two photospheric footpoints. The simulation domain includes the chromosphere, corona and the topmost part of the convection zone at each loop footpoint. The heating of the coronal loop is driven self-consistently by magnetoconvection at the footpoints. The simulation setup is described in detail in \citet{2022A&A...658A..45B}. 
As an initial condition for the magnetic field configuration, we choose a uniform magnetic field strength of 60 G, corresponding to a weak plage region. The simulation was run for 30 min with the new resolution to let initial transients subside before taking data for analysis. We run the simulation at three different grid resolutions, 60
km, 24 km and 12 km. We will refer to the three different runs as LR, MR and HR. The calculation box has a size of $6\times 6\times 57\;$Mm. The convection zone layer has a depth of 3.5 Mm below the photosphere, leading to an effective loop length of 50 Mm. The computational domain is covered by $100\times 100 \times 950$ grid points for the LR simulation, $250\times 250 \times 2375$ grid points for the MR run and $500\times 500 \times 4750$ grid points for the HR run.\\
For run LR, 50 snapshots covering a
time range of one hour were used. For run MR, 25 snapshots
over the range of 40  minutes were used and for run HR 11
snapshots over 25 minutes.\\
The simulation setup is illustrated in Fig. \ref{fig:loop_model}. The figure shows the distribution of plasma with a temperature between 1 MK and 3 MK in the coronal part of the loop and the distribution of the vertical photospheric magnetic field at time 22.21 min for a resolution of 12 km, where t=0 is the start of the time range from which data was taken. 
In the coronal part of the simulation domain, the radiative losses are modelled using an optically thin loss function, while radiative transfer in the grey approximation in local thermodynamic equilibrium (LTE) in the photosphere and chromosphere is used. 
The run with the highest resolution of 12 km is, to our knowledge, the highest resolution simulation of a coronal loop including magnetoconvection to date. 
Since we expect the most accurate results for the non-thermal line widths for the HR run, we use snapshots from this run to compute the distribution of the line widths.
Due to the large size of the output data for the MR and HR runs and the associated computational costs, we only have a high cadence time series with an output cadence of 5 s covering an hour of solar time for the low resolution setup. We therefore use the LR run to follow the evolution of plasma quantities in time. The MR run serves as comparison for the investigation of the dependence of non-thermal line widths on grid resolution.

\subsection{Synthesizing line profiles}
\label{section:synthesis}

In a first step, we compute the emissivity at each gridpoint assuming ionization equilibrium. We synthesize the emission for the \fexv\ 284.163 \AA\ and \fexii\ 195.119 \AA\ lines for each analyzed snapshot. These lines were chosen in order to compare the synthetic line profiles to observations, since both lines are commonly used in solar EUV spectroscopy (see e.g. \citet{2007SoPh..243...19C,2008ApJ...678L..67H,2016ApJ...827...99T}).
The synthesized lines have formation temperatures of about ${\log}_{10}T~[{\rm{K}}]=6.34$ and ${\log}_{10}T~[{\rm{K}}]=6.19$, respectively. The line formation temperatures were taken from the CHIANTI atomic database, version 10 \citep{1997A&AS..125..149D,2021ApJ...909...38D}. The plasma in the coronal part of the loop has an average temperature around 2 MK and is thus expected to be bright in \fexv.
The broadening arising from motions of cooler plasma, especially close to the footpoints, is expected to be captured by the \fexii\ line.\\
Our calculation of the spectral line profiles follows \citet{2006ApJ...638.1086P}. The emissivity in a given gridpoint is given by
\begin{equation}
    \varepsilon = G(T, n_{e})\cdot n_{e}^{2},
\end{equation}
with $T$ and $n_{e}$ being the temperature and electron density and $G(T, n_{e})$ the contribution function for the respective spectral line. The contribution functions were taken from CHIANTI.

To compute the spectral line profiles, we assign a Gaussian line profile to each gridpoint. 
The width of the Gaussian is given by the thermal width, 
\begin{equation}
    w_{\rm{th}}=\sqrt{\frac{2 k_{\rm{B}}T}{m_{\rm{Fe}}}}\label{eq:thermw},
\end{equation}   
arising from the thermal motion of the particles in the plasma.
 The thermal width is computed from the temperature $T$ in each grid cell and the mass of the Fe ion $m_{\rm{Fe}}$ under consideration, with $k_{B}$ being the Boltzmann constant. The Gaussian profile is then shifted according to the line-of-sight (LOS) velocity at each gridpoint. The amplitude of the Gaussian is determined by the peak intensity, which is related to the emissivity by $I_{\rm{peak}}=\frac{\varepsilon}{\sqrt{\pi}w_{\rm{th}}}$.
 The shape of the Gaussian profile in units of Doppler velocity is then given by 
 \begin{equation}
     I_{v}=\frac{\varepsilon}{\sqrt{\pi}w_{\rm{th}}}\exp{\left(- \frac{(v-v_{0})^{2}}{w_{\rm{th}}^2}\right)},
 \end{equation}
 with $v_{0}$ being the velocity component along the LOS.
 Subsequently, we integrate the line profiles along the LOS. For an equidistant grid, we approximate the integral as a sum over the line profiles from each gridpoint along the LOS. The spectra can be calculated in this way because the coronal lines under consideration are optically thin and full radiative transfer calculations are not necessary. We choose four different LOS angles, two perpendicular to and two along the magnetic guide field. Doppler shift and line width are calculated from the first and second moments of the line profile.
 We assume that the simulated fluxtube is isolated, meaning there are no other loop structures in front or behind it when regarding a LOS perpendicular to the magnetic guide field.
 For the LOS parallel to the magnetic guide field, we integrate only up to a distance of 10 Mm above the photosphere along the flux tube, to take into account the effect of looking down at the loop footpoints through the leg of a curved loop. 
\subsection{Non-thermal velocities}
\label{section:non-thermal}

We compute the observed line width $w_{\rm{obs}}$ by calculating the second moment of the line profile, thus mimicking an observation. We do not, however, consider the effects of instrumental broadening.
In order to obtain the non-thermal line width $w_{\rm{nth}}$, we need to subtract the thermal width:
\begin{equation}
    w_{\rm{nth}}=\sqrt{w_{\rm{obs}}^{2}-w_{\rm{th}}^{2}}.\label{eq:width}
\end{equation}
We assume a thermal width $w_{\rm{th}}$ computed using the peak formation temperature of the respective emission line. This ensures that the non-thermal line width is calculated the same way as for observations. One has to remember, however, that the coronal part of the simulation domain does not have a homogeneous temperature. Instead, a large range of temperatures can be present along the line of sight.\\
In regions with temperatures along the LOS below the peak formation temperature, the width of the synthetic spectral profiles  would be smaller than the thermal width as defined in Eq. \ref{eq:thermw} when computed using the peak formation temperature. The term under the square root in Eq. \ref{eq:width} would then be negative.
In real observations, emission in these coronal lines would likely be too small to be detected. We still find some rays which have a low temperature along the LOS, but do have relatively high intensities due to high densities. For the LOS chosen perpendicular to the magnetic guide field, we are looking across the simulation box along the x- or y-direction as illustrated with the coordinate system in Fig. \ref{fig:loop_model}. In real observations it is unlikely that we would look from the side at the cooler footpoint region of a loop, instead there would likely be other structures along the LOS contributing to hot emission, or we would look at the footpoints through hotter plasma higher up in the loop. We therefore discard all rays in the analysis that do not lead to real non-thermal line widths in addition to implementing an intensity threshold.\\ 
Due to limited instrument sensitivity, in an actual observation only the brightest regions would be selected. To exclude regions with low emission, we compute the time-averaged intensity for each line for all the snapshots used in the analysis and exclude rays where the intensity is below 20 \% of the mean intensity for the time series. For calculating histograms of the non-thermal line width, the contributions from different rays are weighted by the intensity of each line profile resulting from the LOS integration.\\
In addition to the distribution of the non-thermal velocities over the simulation domain, we are also interested in the resulting line width for different fields of view, corresponding to different instrument resolutions.
For simplicity, we degrade the synthetic data spatially by using a simple spatial rebinning. In principle, the data should be convoluted with the spatial point spread function of an instrument and then binned to the spatial pixel scale. However, for our purposes this simple rebinning is sufficient. We take the line profile for each ray computed by integrating along the LOS and rebin by different factors by summing up the line profiles from neighbouring rays covering the chosen resolution element. The resulting line width is then calculated as the second moment of the rebinned line profile. We also construct loop-averaged line profiles for a viewing window covering the entire coronal part of the loop for the low-resolution timeseries. 
We perform a comparison between the non-thermal line broadening computed assuming a constant thermal width and the non-thermal width using the actual plasma temperature in Sect. \ref{section:heat}.
\section{Results}
\label{section:res_ntlw}

\begin{figure*}
\resizebox{\hsize}{!}{\includegraphics{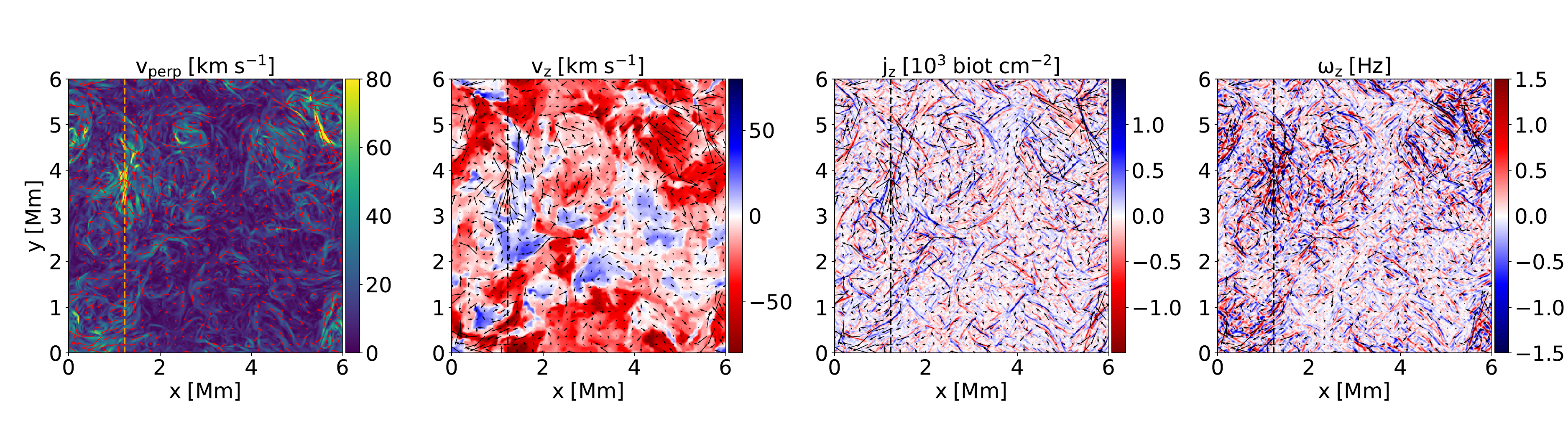}}
  \caption{Cross-section through the loop apex for run HR with $\Delta x = 12\; \rm{km}$ at time 22.21 min. From left to right: Unsigned velocity perpendicular to the loop axis, velocity parallel to the guide field, current density parallel to the guide field, and axial component of the vorticity. The vertical (orange) black line marks the LOS along which the line profile shown in Sect. \ref{section:line_prof} is integrated. The arrows illustrate direction and magnitude of the transverse velocity field. See Sect. \ref{section:res_ntlw}.}
  \label{fig:apex_cs}
\end{figure*}

\begin{figure*}
\resizebox{\hsize}{!}{\includegraphics{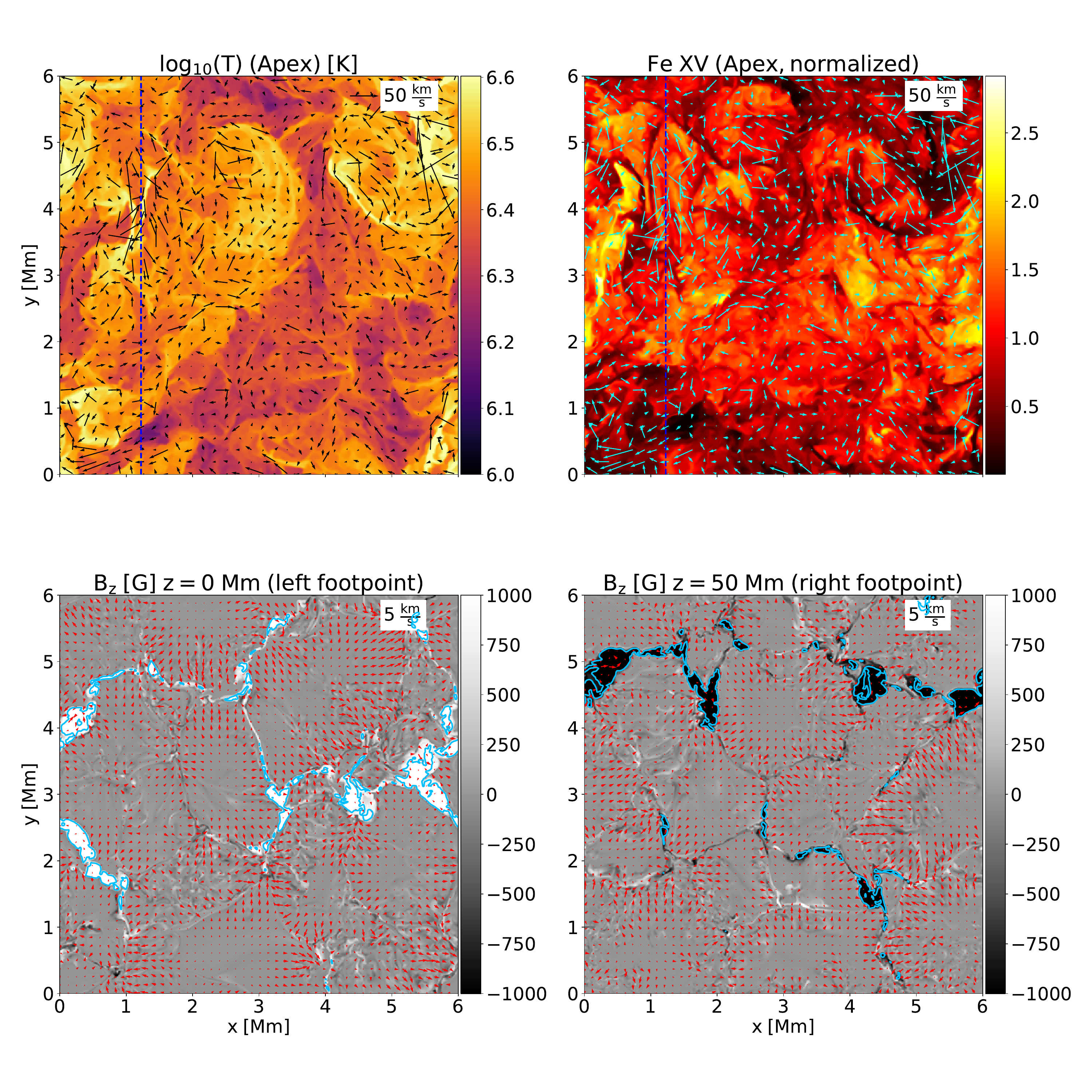}}
  \caption{Overview over the simulation box for run HR with $\Delta x = 12\; \rm{km}$ at time 22.21 min. The top row shows the same cut as Fig. \ref{fig:apex_cs}. Top row: Temperature (left) and emission in the \fexv\ line (right) at the loop apex. The vertical blue line marks the ray along which the line profile shown in the left panel of Fig. \ref{fig:line_profile_asym} is computed. Bottom row: Vertical magnetic field at the solar surface. The blue contours mark the outlines of kilogauss concentrations with $\vert B_{z}\vert > 1000\; \rm{G}$. The arrows illustrate direction and magnitude of the transverse velocity field. See Sect. \ref{section:res_ntlw}.}
  \label{fig:box_overview_ntlw}
\end{figure*}

In response to the photospheric driving, turbulent-like behaviour develops in the coronal loop. Multiple small current sheets form at length scales below the resolution limit of current instruments. A cross section of velocities, current density and vorticity at the loop apex is shown in Fig. \ref{fig:apex_cs}. The transverse velocity components show a complex small-scale structure, leading to high vorticity, while the velocity parallel to the guide field is organized on larger length scales. The magnetic field distribution at the loop footpoints and the distribution of temperature and emission in the \fexv\ line at the apex are shown in Fig. \ref{fig:box_overview_ntlw}. The temperature in the loop cross section ranges from 1.6 MK to 7.3 MK. Since a significant amount of plasma is above the line formation temperature and the emission depends quadratically on the density, the distribution of the emission in \fexv\ deviates from the temperature distribution. It also shows that over only 6 Mm across the loop, a large variation in temperature can occur. This has consequences for the interpretation of the obtained spectra.
Despite the large range in plasma temperature, we chose not to use an emission line that forms at higher temperatures, since the filling factor of very hot plasma with temperatures of $\log(T)>6.5 $, at which the response function of the \fexv\ line has dropped to half its peak value, is small, being roughly 8 \% in the snapshot shown in Fig. \ref{fig:box_overview_ntlw}. In the snapshots used for the analysis in Sect. \ref{section:perp} and \ref{section:par}, the filling factor is below 8 \%.\\
In the following sections, we discuss the properties of the line profiles obtained from forward modelling, their dependence on the spatial resolution of the synthetic observations as well as their connection to heating events and resulting plasma flows.
\subsection{Non-thermal broadening perpendicular to guide field}
\label{section:perp}

\begin{figure*}
\resizebox{\hsize}{!}{\includegraphics{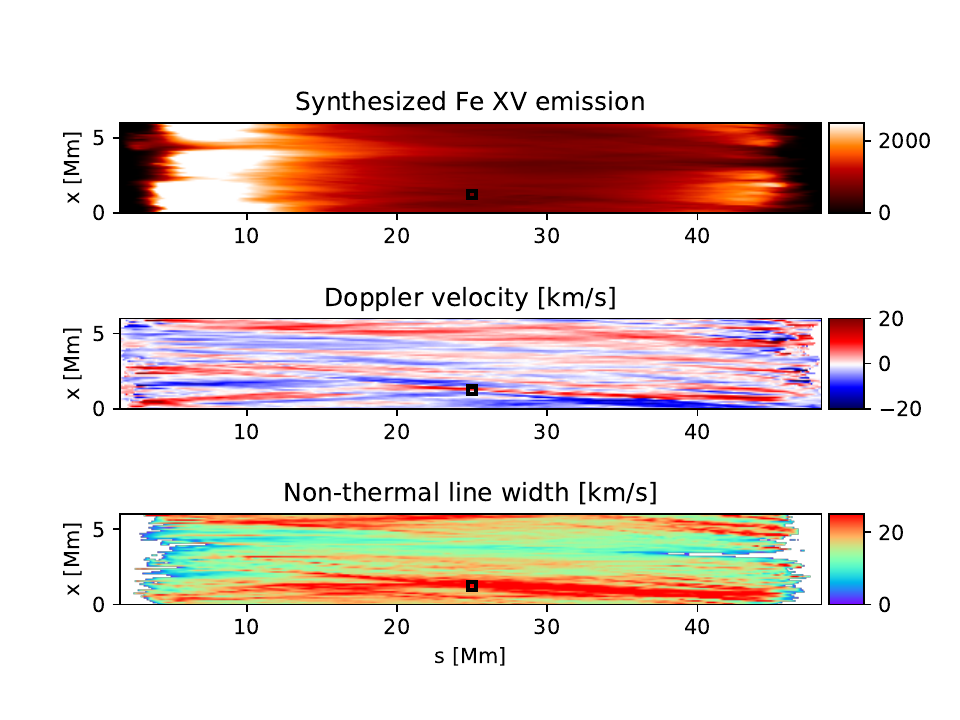}}
  \caption{Spatial maps in intensity, Doppler shift and line width. From top to bottom: Intensity of the emission in the \fexv\ line, Doppler shift, and non-thermal line broadening seen perpendicular to the guide field for run HR at time 22.21 min. The line of sight integration was performed along the y-axis. We mask regions with low emission that lead to line widths below the thermal width computed from the peak formation temperature in white. The black square marks the field of view used to compute the line profile in Fig. \ref{fig:spec_vel}. The line profile is integrated over the area of the square. The integrated emission has the units $\rm{[erg\; cm^{-2}s^{-1}sr^{-1}]}$. For a discussion see Sect. \ref{section:perp}.}
  \label{fig:ntlw_hr}
\end{figure*}

\begin{figure*}
\resizebox{\hsize}{!}{\includegraphics{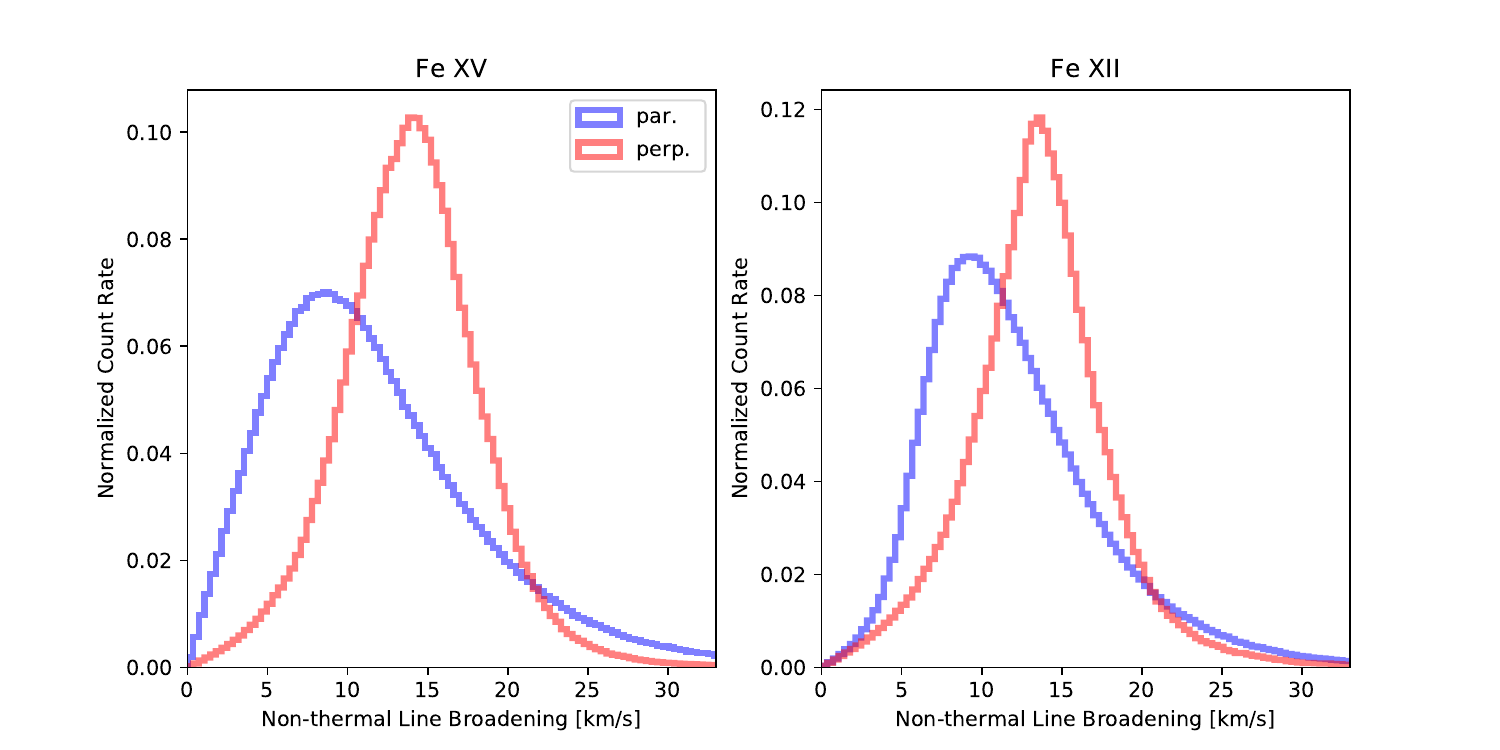}}
  \caption{Time-averaged normalized histograms for the non-thermal line width in \fexv\ (left panel) and \fexii\ (right panel) seen perpendicular to the guide field (red) and parallel to the guide field (blue). We show the histograms for the highest grid resolution of 12 $\rm{km}$. The distributions are weighted with the intensity. See Sect. \ref{section:perp} and Sect. \ref{section:par}.}
  \label{fig:hist_Fe15_LOS}
\end{figure*}

The maps obtained from a LOS-integration along the x- and y-direction correspond to a loop seen from the side above the limb, with only motions transverse to the magnetic field contributing to the line broadening.
LOS-integrated emission, Doppler shift and non-thermal broadening for run HR are shown in Fig. \ref{fig:ntlw_hr} for an integration along the y-direction. Emission, Doppler shift and non-thermal line broadening exhibit elongated structures aligned with the guide field. The emission is strongest close to the footpoints despite the higher temperature near the apex due to the higher density in the low corona. At the apex, the average temperature is 2.7 MK, lying above the \fexv\ line formation temperature of roughly 2.2 MK. The brightest emission is therefore found in the loop legs instead of the apex. The left loop leg is brighter than the right leg, as both temperature and density are higher in the left loop leg for the snapshot shown, but this is changing with time. For the \fexii\ line, the emission is even more strongly concentrated towards the footpoints and lower at the loop top. \\
The brightest regions do not coincide with the largest line widths or Doppler shifts. Near the loop apex, the highest Doppler shifts and values for the line broadening appear in the dark regions. The correlation between non-thermal line widths and intensity is very weak or even moderately negative for some snapshots for \fexv\ and moderately negative for \fexii\ for run HR. The hottest plasma in the loop has a temperature of 4.3 MK to 12.5 MK for the analyzed snapshots and thus does not appear bright in \fexv. At these temperatures, the contribution function for this line has fallen to about 8 \% and less than one percent of its maximum value.
While areas showing a large Doppler shift also exhibit increased non-thermal broadening, the correlation between the magnitude of the Doppler shift and the non-thermal line broadening is weak for both \fexv\ and \fexii.\\
Fine strands are not only present in the emission, but also in the Doppler shift and non-thermal line broadening. Various parts of the loop seem to move independently.\\
The statistical properties of the non-thermal line broadening as would be measured perpendicular and parallel to the magnetic guide field are shown in  Fig. \ref{fig:hist_Fe15_LOS} for the \fexv\ and \fexii\ lines.
Since the non-thermal line broadening varies with changing temperature and velocities in the loop, we have computed normalized intensity-weighted histograms for a range of snapshots and then averaged the histograms in time. Data for line-of-sight integration in both the x- and the y-direction was included in the distribution. 
Areas with a line width smaller than the thermal width computed for the peak formation temperature of the emission line were excluded.
For the line-of-sight perpendicular to the magnetic guide field, the peak and mean values of the distribution for the \fexv\ line are the same with 13.8 $\rm{km\; s^{-1}}$. For the \fexii\ line, the peak and the mean value are 13.4 $\rm{km\; s^{-1}}$. 

\subsection{Non-thermal broadening parallel to guide field}
\label{section:par}

\begin{figure*}
\resizebox{\hsize}{!}
{\includegraphics{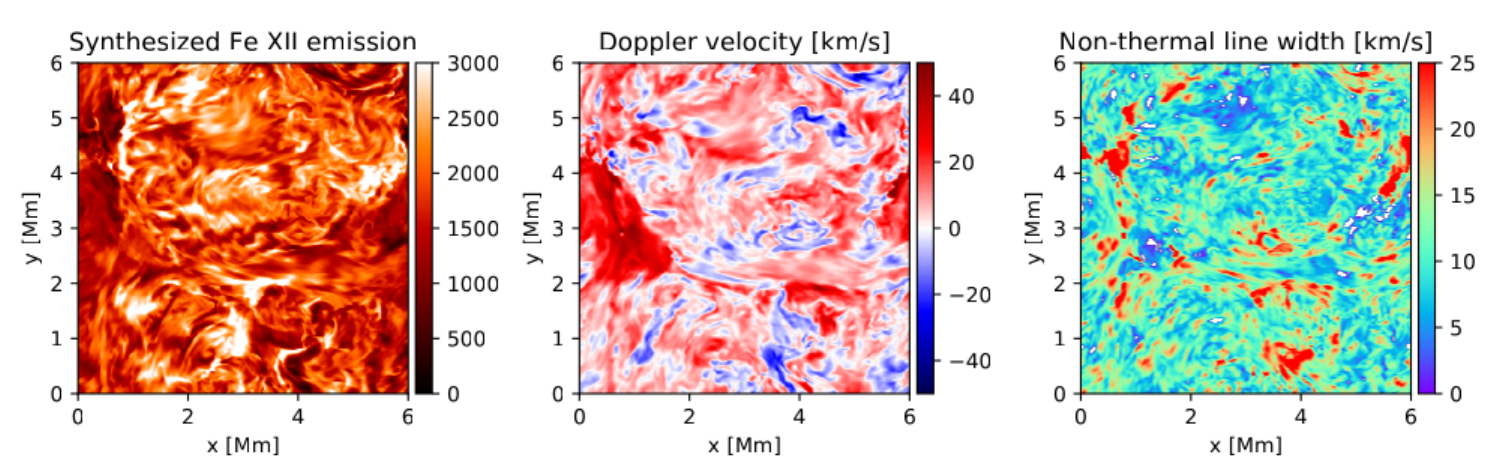}}
  \caption{From left to right: Intensity of the emission in the \fexii\ line integrated along the line of sight over the range s $\rm{\epsilon}$ [0,10] Mm, Doppler shift and non-thermal line broadening seen parallel to the guide field for the left loop footpoint and run HR. The integrated emission has the units $\rm{[erg\; cm^{-2}s^{-1}sr^{-1}]}$. For a discussion, see Sect. \ref{section:par}.}
  \label{fig:wnth_fp_12}
\end{figure*}

In order to simulate an observation of the loop footpoints close to disk center, we integrate the line profiles along a LOS parallel to the guide field.
Since an observer would not look down from the apex along the curved magnetic field to the footpoints, but only see part of the loop leg, we limit the range of integration to $s\;\epsilon\; [0,10]\; \rm{Mm}$. This captures roughly the footpoint region of the 50 Mm long semi-circular loop that an observer would see at disk center.\\
The emission in the \fexii\ line, Doppler shift and non-thermal broadening for the left loop footpoint are shown in Fig. \ref{fig:wnth_fp_12}. 
The non-thermal broadening is organized in patches.\\ 
The distribution of the non-thermal broadening is shown in Fig. \ref{fig:hist_Fe15_LOS}  for the \fexv\ and \fexii\ emission. Peak and average values of the distribution of non-thermal line widths are slightly lower than for the view perpendicular to the guide field.

Seen parallel to the magnetic guide field, the average non-thermal line broadening at the footpoints in the \fexv\ line is 11.5 $\rm{km\; s^{-1}}$ and the peak is at 8.5 $\rm{km\; s^{-1}}$ in the \fexv\ emission. These values are roughly similar for the \fexii\ emission with a peak value of 9.2 $\rm{km\; s^{-1}}$ and a mean of 11.9 $\rm{km\; s^{-1}}$.
 Similar to the LOS perpendicular to the guide field, the correlation between non-thermal broadening and Doppler shift is weak for both the \fexv\ and the \fexii\ line. The correlation between non-thermal velocities and intensity is negligible, not negative as in the perpendicular case.
 Despite showing similar mean values, the distributions of the non-thermal line broadening perpendicular and parallel to the loop axis have different shapes.
 In contrast to the distribution of the non-thermal line widths seen perpendicular to the loop axis, the distribution of the non-thermal line widths computed for the direction of the line of sight parallel to the loop axis exhibits a highly skewed distribution.

\subsection{Non-thermal broadening and heating}
\label{section:heat}

\begin{figure}
\resizebox{\hsize}{!}{\includegraphics{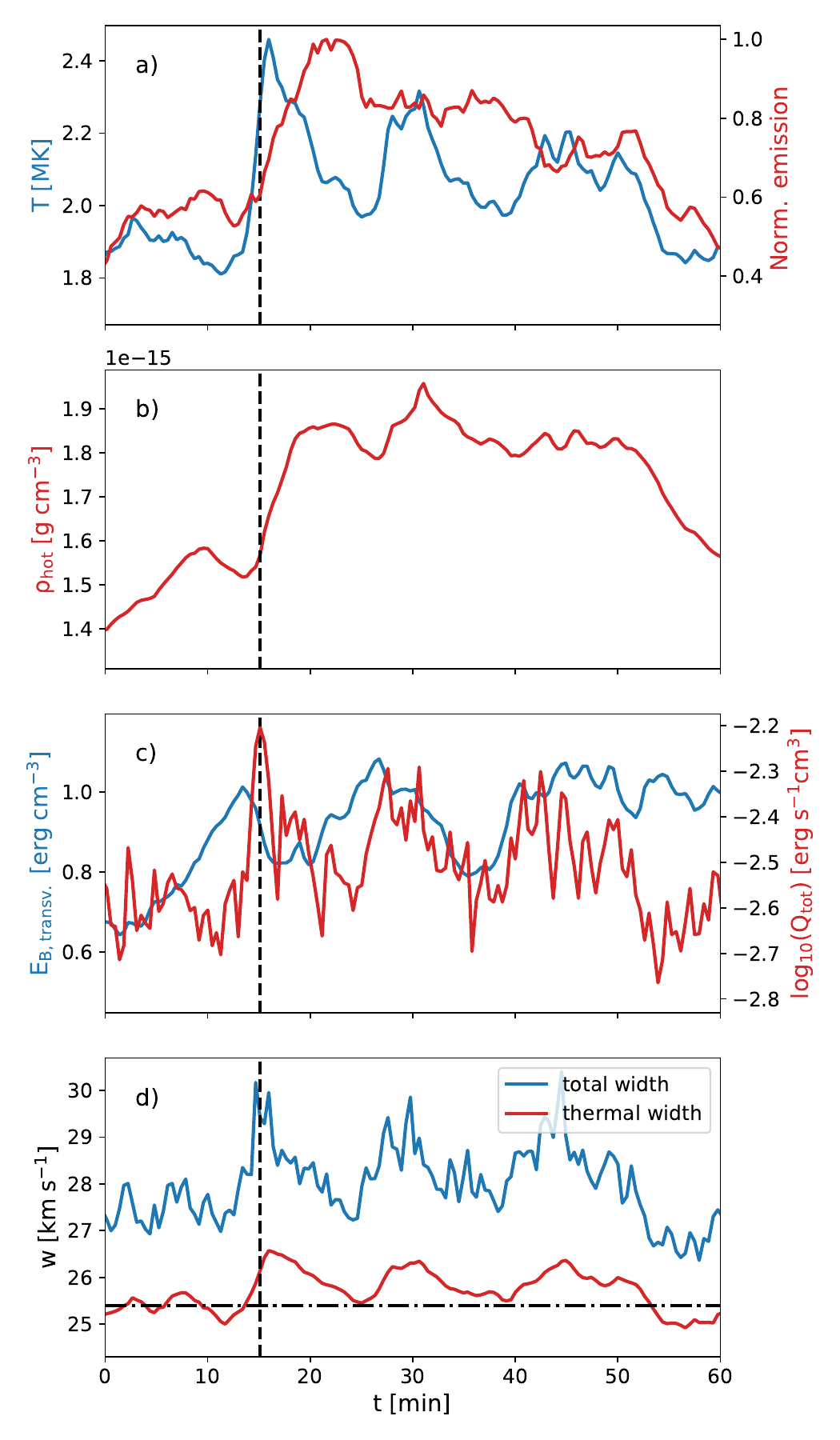}}
  \caption{From heating to increased line width. Time evolution of the coronal loop for run LR. Panel (a) shows the temperature (blue) and normalized emission in the \fexv\ line (red) averaged over the coronal part of the loop, defined as grid points with a mass density below $10^{-12}\; \rm{g\; cm^{-3}}$. The average density for hot plasma between 1.5 MK and 2.9 MK is shown in panel (b). Panel (c) shows the sum of viscous and resistive heating rate (red) and the magnetic energy density computed from the transverse components of the magnetic field (blue), and panel (d) shows the total spectral line width (blue) measured for a LOS along the y-axis and a field of view corresponding to the whole coronal part of the simulation box from s=1.5 Mm to 48.5 Mm. The solid red line shows the evolution of the thermal line width integrated over the same region. The vertical dashed line marks the time of the strongest heating event. The dash-dotted horizontal line denotes the thermal line width of 25.4 $\rm{km\; s^{-1}}$ at the peak formation temperature of \fexv\ of 2.2 MK. See Sect. \ref{section:heat}.}
  \label{fig:time_evo}
\end{figure}

\begin{figure*}
\resizebox{\hsize}{!}{\includegraphics{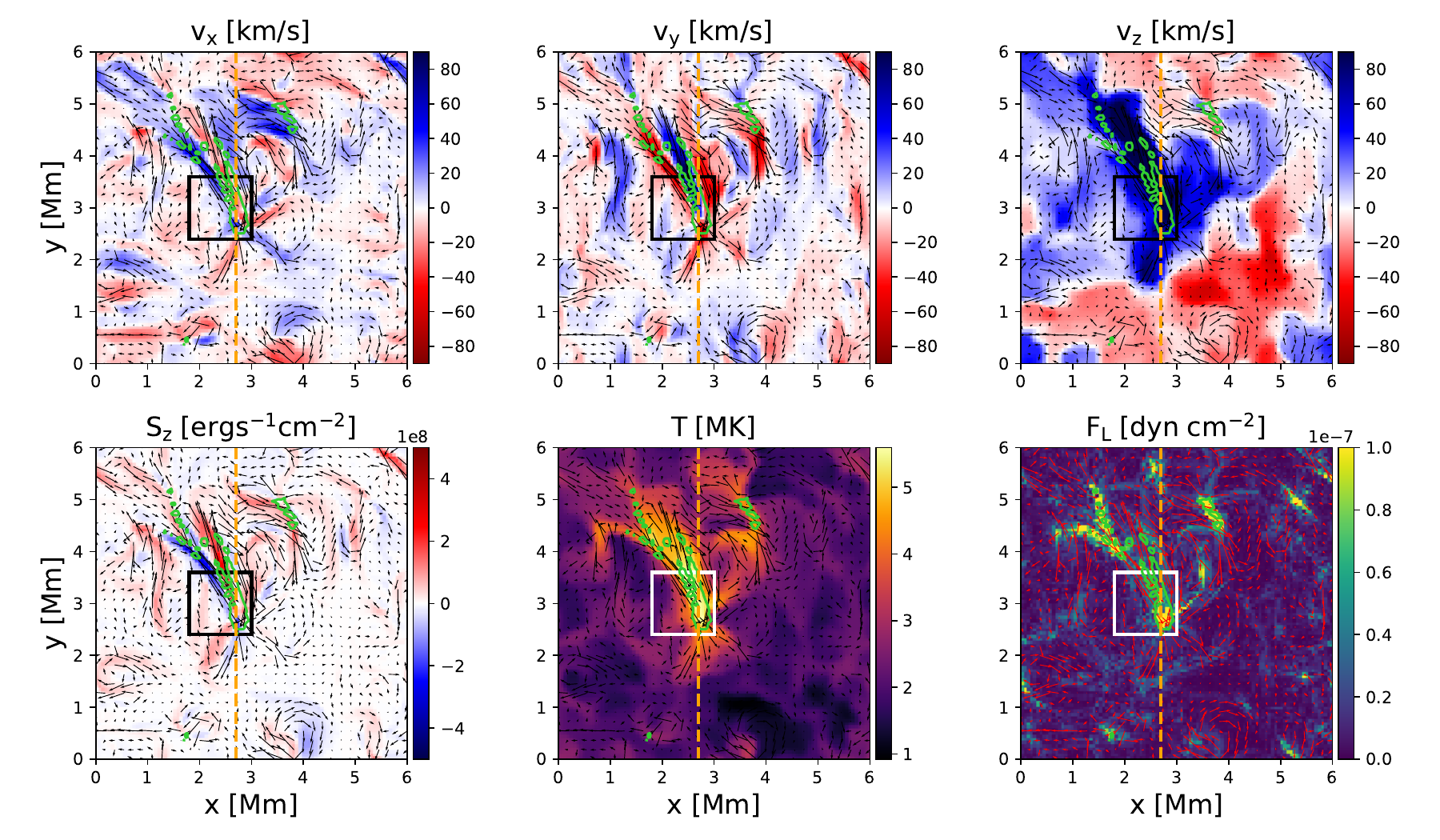}}
  \caption{Cross-section through the strongest heating event in the timeseries shown in Fig. \ref{fig:time_evo} at time 17.9 min and a height of 11.26 Mm. Top row from left to right: Transverse velocity components $v_{x}$, $v_{y}$ and field-aligned velocity $v_{z}$. Bottom row from left to right: Axial component of the Poynting flux, temperature, and transverse component of the Lorentz force. The green contour outlines the regions with a total heating rate above a threshold value of $0.1\; \rm{erg\; s^{-1} cm^{-3}}$. The black arrows illustrate direction and magnitude of the transverse velocity field. The white and black squares mark the location of the $1.2\times 1.2 \times 1.2$ Mm box containing the strongest heating event determined as described in \ref{section:heat_ind}.The vertical orange line marks the ray along
which the line profile shown in Fig. \ref{fig:line_profile_lr} is computed. For a discussion see Sect. \ref{section:heat}.}
  \label{fig:vel_comp}
\end{figure*}

\begin{figure}
\resizebox{\hsize}{!}{\includegraphics{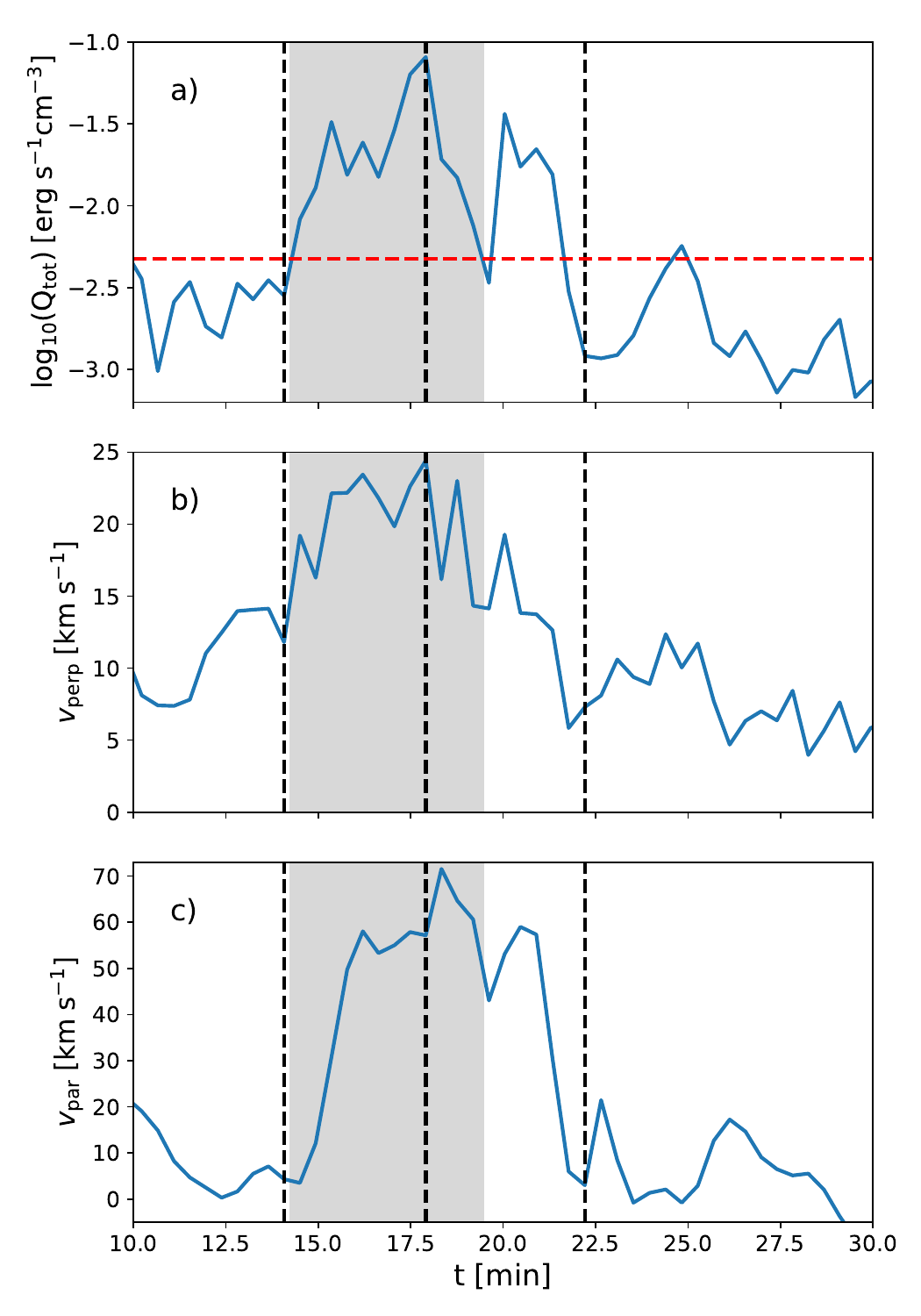}}
  \caption{Time evolution of the total heating rate (a) and perpendicular (b) and parallel velocity components averaged over a $1.2\times 1.2\times 1.2\; \rm{Mm}$ box containing the strongest heating event at 17.9 min. The red dashed line marks the time-averaged heating rate for the entire time series. The shaded area corresponds to the strongest peak in heating. The black dashed lines mark the times of the snapshots shown in Fig. \ref{fig:braid}. See Sect. \ref{section:heat}.}
  \label{fig:time_evo_cube}
\end{figure}

\begin{figure}
\resizebox{\hsize}{!}{\includegraphics{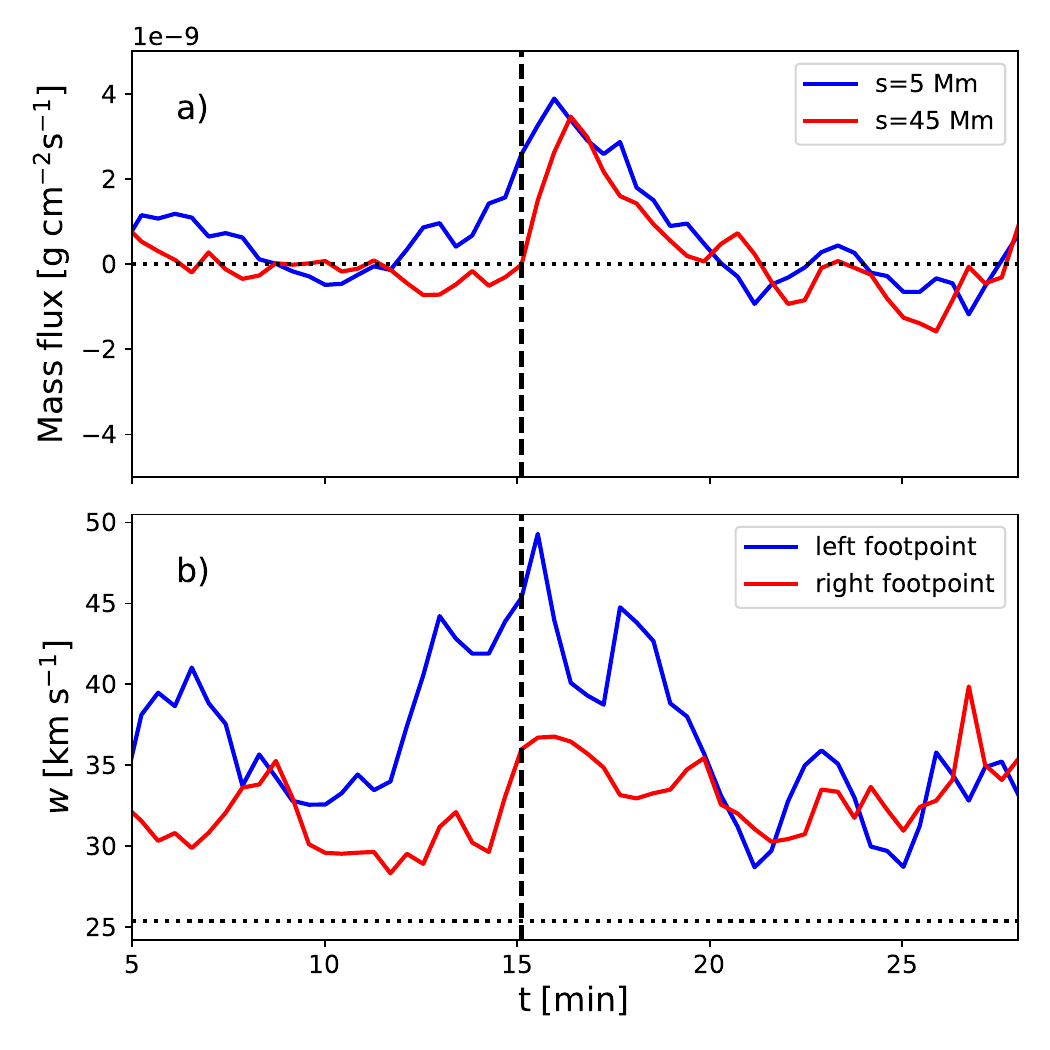}}
  \caption{Time evolution of mean mass flux through a cross-section at height 5 Mm above each loop footpoint (panel (a)) and spectral line width (panel (b)) synthesized with a line-of-sight parallel to the guide field for the emission in the \fexv\ line integrated over the loop cross-section for each footpoint. The vertical dashed line marks the time of the strongest heating event (see Fig. \ref{fig:time_evo}), while the dotted horizontal line denotes the zero line in panel (a) and the thermal line width of 25.4 $\rm{km}$ at the peak formation temperature of \fexv\ of 2.2 MK. See Sect. \ref{section:heat}.}
  \label{fig:time_evo_par}
\end{figure}

\begin{figure*}
\resizebox{\hsize}{!}{\includegraphics{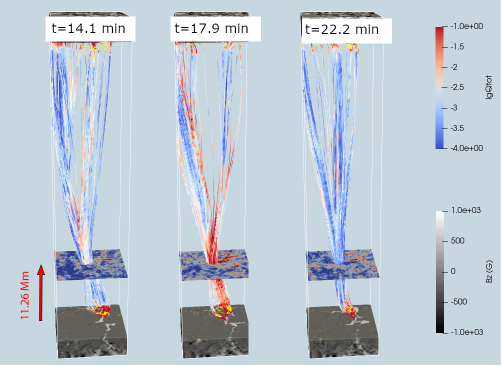}}
  \caption{Magnetic field lines shortly before, during and after the strongest heating event for run LR. The time stamp of the snapshots is marked by black vertical lines in Fig. \ref{fig:time_evo_cube}. The shading on the plane at 11.26 Mm and the coloring of the field lines show the logarithm of the sum of the viscous and resistive heating rate.  The axial direction has been compressed by a factor of two for better visibility.} 
  \label{fig:braid}
\end{figure*}

\subsubsection{Time evolution}

The non-thermal line width varies in time. We constructed a timeseries of the loop-averaged non-thermal broadening for the $\Delta x=60\; \rm{km}$ resolution run.  The time evolution of the coronal averages of various quantities is shown in Fig. \ref{fig:time_evo}. The coronal part of the simulation domain is here defined as the region with a mass density below $10^{-12}\; \rm{g\; cm^{-3}}$. The temperature shows several peaks, with the strongest event at 17.9 min, and subsequent cooling. The emission increases in response to the increased temperature. The peaks in the emission show a delay of several minutes with respect to the temperature peaks. This is due to two effects. First, the maximum average temperature reached in the simulation is above the line formation temperature of \fexv\, so the emission is expected to peak when the loop cools. Due to the strong density dependence of the optically thin emission, the emission increases in response to the density increase by chromospheric evaporation following a heating event. The emission reaches a peak when the average density of hot plama between 1.5 MK and 2.9 MK reaches its first maximum after the strongest heating event (see panel (b) of Fig. \ref{fig:time_evo}). Therefore, the maxima in the emission occur after the maxima in the average temperature (see also the discussion in \citet{2022A&A...658A..45B}). \\
All the temperature maxima are preceded by peaks in the heating rate. The second panel shows the time evolution of the heating rate.  The strongest peak is located at time 17.9 min. 
The blue curve is the magnetic energy density arising from the transverse field components alone. For the stratified loop setup with an initially uniform, axial magnetic field that we use in our simulation, the potential magnetic field is close to axial in the coronal part of the loop. Thus the magnetic energy density associated with the transverse field components is an approximation for the available free magnetic energy. The magnetic energy density drops shortly before the first and second strong heating event. The third heating event is associated with an initial increase in magnetic energy density, followed by several smaller dips.\\
The time development of the line width seen perpendicular to the loop axis follows the time evolution of the heating rate, with the largest line width associated with the strongest peak in the heating rate. A correlation between the temperature and the velocity dispersion was also found in \citet{2016A&A...589A.104G}.
Panel (d) of Fig. \ref{fig:time_evo} shows the spectral line width for a viewing window covering most of the length of the loop between s=1.5 Mm 
and 48.5 Mm as a function of time.
To check whether the peaks in the line broadening arise from increased velocity fluctuations in the loop or from increased plasma temperature, we compute the line profiles for the timeseries with the Doppler shift set to zero. Panel (d) of Fig. \ref{fig:time_evo} shows the line width that arises from thermal broadening alone. While the evolution of the thermal line width roughly follows the evolution of the temperature, the thermal width depends on the square root of the temperature. Even for a steep increase in temperature, the thermal width increases only by a maximum of 1.5 $\rm{km\; s^{-1}}$ above the thermal width at the line formation temperature. We find that the effect from the small-scale motions dominates and the increase in non-thermal line broadening still persists after subtracting the thermal line width. The line width always exceeds the thermal width during the simulation time.\\
\subsubsection{Behavior of an individual heating event}
\label{section:heat_ind}
To study the relation between non-thermal broadening, the velocity field and heating events in more detail, we take a closer look at the origin of the strongest peak in the heating rate for the LR run. We divide the simulation domain into $1.2\times1.2\times 1.2\; \rm{Mm}$ subdomains and determine the cube containing the strongest heating event. The cube containing the strongest heating is centered on the point [x,y,z]=[2.4,3,11.26] Mm.
A cross-section through the strongest heating event is shown in Fig. \ref{fig:vel_comp}. The green contour outlines the location of the heating event, with a threshold of $Q_{\rm{tot}}=0.1\; \rm{erg\; s^{-1} cm^{-3}}$. Axial Poynting flux, viscous and resistive heating rates are increased, with viscous heating being the dominant type of heating. The main channel of dissipation is therefore not Joule heating, but thermalization of the reconnection outflows. In response to the heating, temperatures exceeding five MK are achieved. The heating event is located over a strong gradient in the $v_{x}$- and $v_{y}$- components of the velocity, for example at the location of a strong shear flow which can be seen in Fig. \ref{fig:vel_comp} in the region marked with a square. The event occurs at the location of several misaligned  magnetic field strands as illustrated in Fig. \ref{fig:braid}. 
The energy content of a box with dimensions $1.2\times 1.2\times 15\; \rm{Mm}$ integrated over a time of 332 s covering the strongest spike in heating (shaded area in Fig. \ref{fig:time_evo_cube}) is $2.6\times 10^{25}\;\rm{erg}$, a value compatible with the nanoflare energy of $10^{24}\; \rm{erg}$ to $10^{27}\; \rm{erg}$ for the strongest events suggested by \citet{1988ApJ...330..474P}. While the energy content is higher than the $10^{23}-10^{24}\; \rm{erg}$ typical for a nanoflare, this is the strongest event in our time series and potentially consists of mutiple heating events that are not temporally resolved. The magnetic field topology at the location of the heating event, consisting of several crossing magnetic flux bundles, supports the braiding picture. Impulsive energy releases, however, can also be triggered by waves \citep{2018MNRAS.476.3328M}.\\ 
A strong upward flow is present in the component of the velocity parallel to the loop axis at the location of the heating event. The horizontal component of the Lorentz force is increased at the heating site, leading to the acceleration of flows. This structure is similar to the one  visible in run HR at the loop apex. Due to the lower numerical diffusivity in run HR, the current sheet is narrower, but the qualitative structure of the current sheet is similar in the HR and LR simulations.\\
\subsubsection{Velocity components}
To investigate whether the increase of the different velocity components at the heating site is a cause of or a response to the heating event shown in Fig. \ref{fig:vel_comp}, we plot  in Fig. \ref{fig:time_evo_cube} the time evolution of the total heating rate, the perpendicular and the parallel velocity in the previously determined region of size $1.2\times 1.2\times 1.2\; \rm{Mm}$ for a duration of 20 minutes around the strongest heating event in the time series for the 60 km run. As panel b) in Fig. \ref{fig:time_evo} shows, the average heating rate shows several strong peaks but never falls below about ${\log_{10}}Q_{\rm{tot}}~\rm{[erg\; s^{-1}cm^{-3}]}=-3$. The velocity perpendicular to the guide field rises prior to the steep increase in heating rate, while the rise in the velocity component parallel to the guide field occurs with a delay of about one minute. While the transverse velocity reaches peak values of about $25\; \rm{km\; s^{-1}}$, the parallel flow is much stronger with speeds of over $60\; \rm{km\; s^{-1}}$. The transverse velocity starts to increase shortly before the heating rate and the axial velocity starts to increase after the onset of the heating event. Therefore, it can be assumed that the heating is due to the shear flow perpendicular to the guide field induced by the evolving magnetic field shown in Fig. \ref{fig:vel_comp}. The increase in the axial component is due to an evaporative upflow in reaction to the heating event.\\
The presence of the strong upflow indicates that there should also be an increase in the line width at the loop footpoint closest to the heating event for the LOS parallel to the guide magnetic field.
The time evolution of the non-thermal broadening of the \fexv\ line seen parallel to the guide field integrated over the entire loop cross-section and the mass flow through a slice located in the low corona at 5 Mm above the photosphere are shown in Fig. \ref{fig:time_evo_par}. For both footpoints, the heating event is followed by an increase in the mass flow with a delay of about 30 s. The heating event is also accompanied by a strong peak in the non-thermal line broadening for the footpoint closest to the heating event, while the other footpoint does not show a clear peak. 

\begin{figure}
\resizebox{\hsize}{!}{\includegraphics{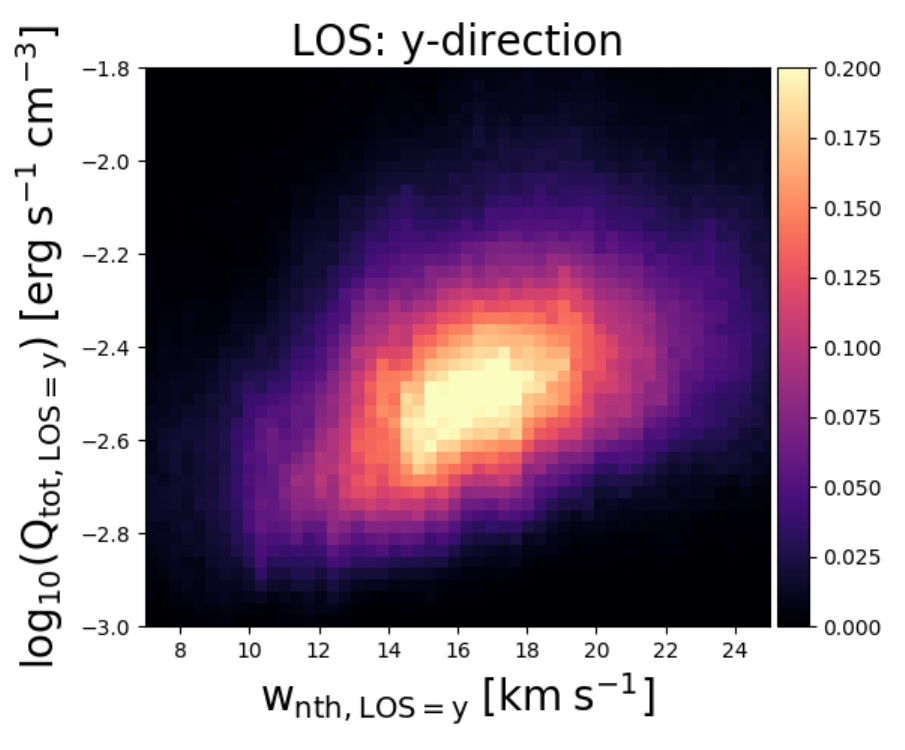}}
  \caption{2D histogram relating the LOS-averaged total heating rate to the non-thermal line broadening for the snapshot shown in Fig. \ref{fig:ntlw_hr}. The non-thermal broadening was calculated for emission in the \fexv\ line. Shown here is the histogram for a LOS along the y-direction. The histogram for a LOS-integration along the x-direction shows similar behaviour. The histogram was computed for the coronal part of the domain between 3 and 47 Mm. See Sect. \ref{section:heat}.}
  \label{fig:heat_lw_histo}
\end{figure}
\subsubsection{Correlation between heating and non-thermal line width}

Since the non-thermal line width shows an increase during a strong heating event, it is interesting to consider whether the line width can be used as a proxy for the heating rate. Here we compute the heating rate as the sum of resistive and viscous heating. Since the viscous heating dominates over the resistive heating in the coronal part, the total heating rate follows the behavior of the viscous heating \citep{2017ApJ...834...10R}.
Using Spearman's rank coefficient $r_{s}$ as a measure of the strength of the correlation, we find that there is a weak to moderate correlation between the non-thermal line width and the LOS-averaged total heating rate for the emission observed perpendicular to the guide field. Since a large fraction of the plasma in the corona is above the formation temperature of the \fexii\ line, we consider here only the correlation between the heating rate and the \fexv\ line.
 A 2D histogram for the relation between non-thermal line broadening of the \fexv\ line and LOS-averaged heating rate is shown in Fig. \ref{fig:heat_lw_histo} for the snapshot of the HR run shown in Fig. \ref{fig:ntlw_hr} at a time of 22.21 min. For this snapshot and choice of LOS, Spearman's rank coefficient is 0.42.\\
A similar moderate correlation exists between the non-thermal line width and the unsigned Poynting flux as well as vorticity. The strength of the correlation varies for different snapshots and choices of the LOS.
A heating event can lead to an increase in line broadening due to either plasma flows or thermal broadening caused by the temperature increase itself. As described in Sec. \ref{section:meth}, when computing the non-thermal line width, we subtract the thermal broadening for the peak formation temperature of the respective emission line according to Eq. \ref{eq:width}.
To check if the correlation between the non-thermal line width and the heating rate is not solely due to an increased thermal width in response to the heating, we isolate the effect of velocity fluctuations on the line width. To this end, we calculate the line profiles for the snapshot shown in Fig. \ref{fig:ntlw_hr} using a fixed thermal line width of 25.4 $\rm{km\; s^{-1}}$ corresponding to the peak formation temperature as the width assigned to the Gaussian profile for each grid point before integrating over the LOS. Subtracting the thermal width then leaves only the contribution from velocity fluctuations. The correlation still holds with Spearman's rank coefficient of 0.38 for a LOS along the x- and y-axis. Hence, the increase in line broadening is mainly due to the (not resolved) flows and not due to the increased temperature.

\subsection{Line profiles}
\label{section:line_prof}

\begin{figure}
\resizebox{\hsize}{!}{\includegraphics{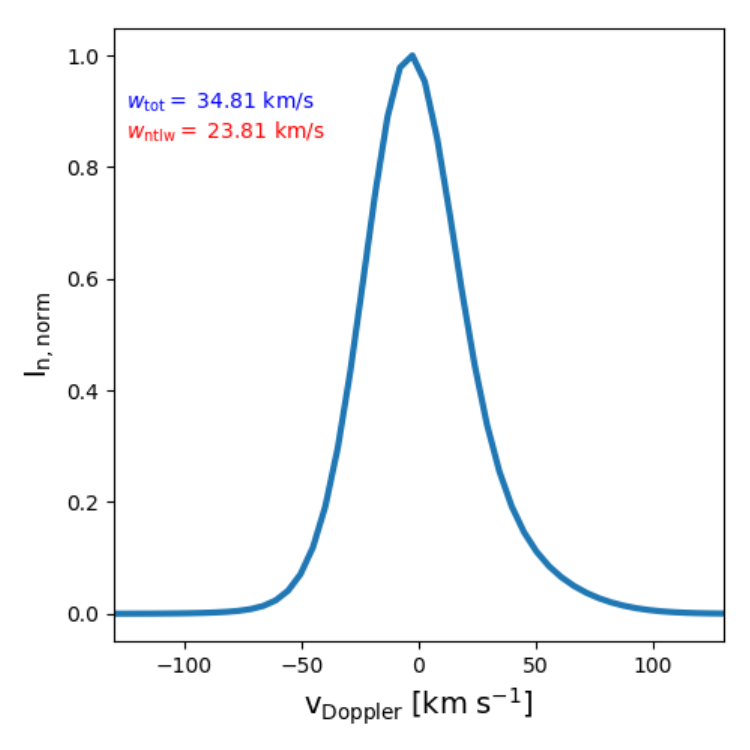}}
  \caption{Synthesized line profile for the \fexv\ line along the ray intersecting the heating event across the guide field shown in Fig. \ref{fig:vel_comp}. The number in blue indicates the total line width and the number in red the non-thermal line width after subtraction of the thermal line width corresponding to the line formation temperature. See Sect. \ref{section:line_prof}.}
  \label{fig:line_profile_lr}
\end{figure}

\begin{figure}
\resizebox{\hsize}{!}{\includegraphics{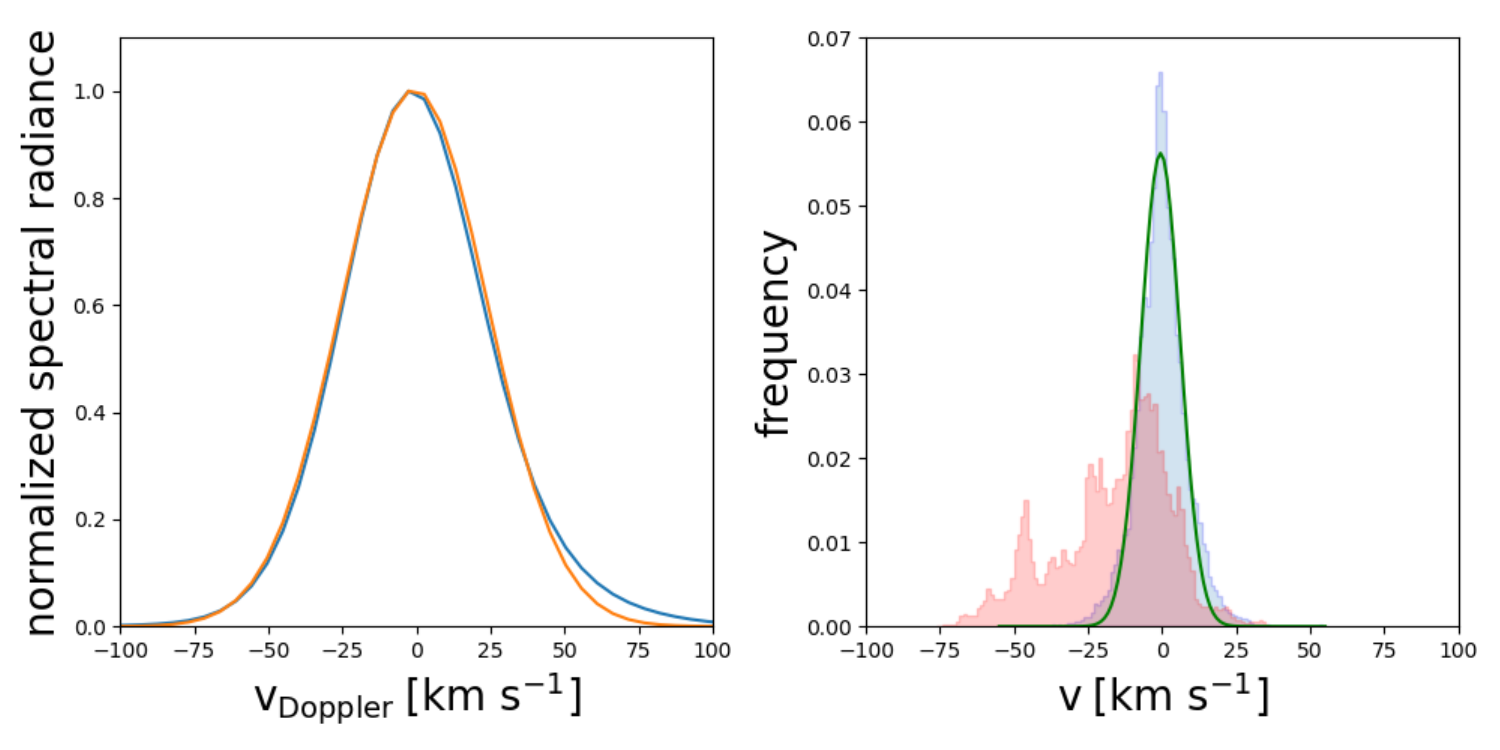}}
  \caption{Synthesized line profile and velocity distribution for a field of view of $0.5\times 0.5\; \rm{Mm} $ at the location highlighted by the black rectangle in Fig. \ref{fig:ntlw_hr}. Left panel: Line profile of \fexv\ with wavelength in Doppler units. The orange line shows a Gaussian fit to the profile. Right panel: Histogram of the LOS-velocity for a LOS in the y-direction (blue) and z-component of the velocity (red) in the field of view. The histograms have been normalized so that the area under the curve is unity. The green line shows a single Gaussian fit to the distribution of the velocity component in the LOS direction. See Sect. \ref{section:line_prof}.}
  \label{fig:spec_vel}
\end{figure}

\begin{figure}
\resizebox{\hsize}{!}{\includegraphics{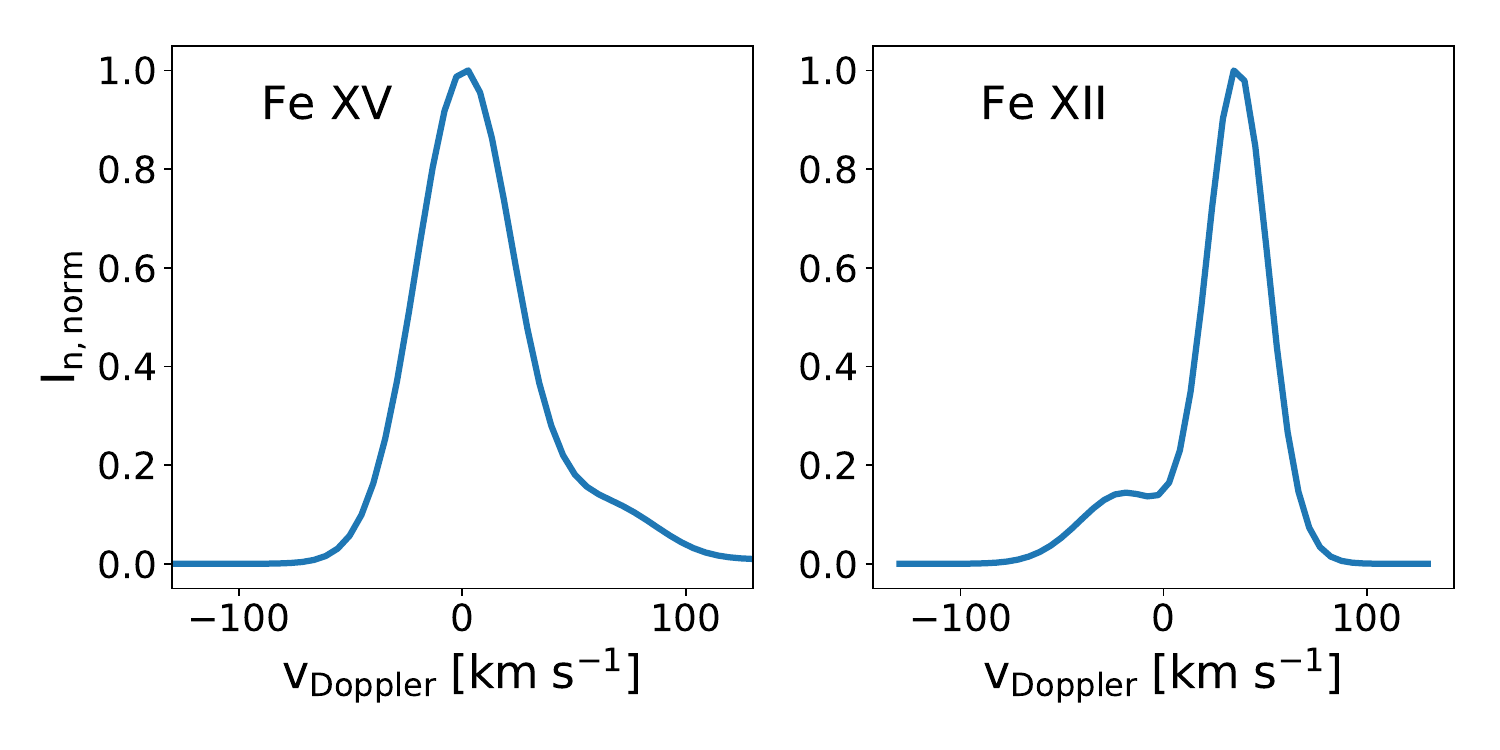}}
  \caption{Examples of asymmetric line profiles. Left panel: Line profile seen in \fexv\ for a LOS along the
y-direction at time 22.21 min for a single grid point. The LOS is shown in Fig. \ref{fig:apex_cs} as a
vertical black line.  Right panel: Example for a line profile with excess emission in the line wings seen in
\fexii\ for a LOS along the guide field for the high resolution run at time 5.2 min for
a single grid point. The normalized spectral radiance is shown as a function of Doppler
velocity. See Sect. \ref{section:line_prof}.}
  \label{fig:line_profile_asym}
\end{figure}

Information about plasma flows can be gained not only from the width of the observed line profiles, but also their shapes.
Line profiles in observations often show a non-Gaussian shape with enhanced line wings \citep{1977ApJ...211..579K, 1993SoPh..144..217D,1997Natur.386..811I, 2008ApJ...678L..67H,2010A&A...521A..51P}. Such non-Gaussian profiles can also be found in simulations \citep{2020A&A...639A..21P}.  Most line profiles in our simulation have a roughly Gaussian shape, but we also find asymmetric profiles with excess emission in the wings. 
In observational studies, these are often interpreted as reconnection jets. In our HR simulation we find that outflows close to 100 $\rm{km\; s^{-1}}$ caused by a strong heating event are present along the ray marked by a dashed line in the top row of Fig. \ref{fig:apex_cs} and Fig. \ref{fig:box_overview_ntlw}. In addition to the outflow, the vorticity is also increased at the location of the heating event. A strong outflow is also present for the strong heating event occuring in the LR run shown in Fig. \ref{fig:vel_comp}. The line profile of \fexv\ arising from integration along a ray passing through the heating event across the guide field is shown in Fig. \ref{fig:line_profile_lr}. The profile has a non-thermal width of 23.81 $\rm{km\; s^{-1}}$, far above the mean value for the LR run of 10.3 $\rm{km\; s^{-1}}$, and has an asymmetric shape with an enhanced line wing.\\
For the HR run, the resulting asymmetric line profile arising from a ray crossing a strong heating event has a broad minor component as is shown in the left panel of Fig. \ref{fig:line_profile_asym}. An example of an asymmetric profile seen along a line of sight parallel to the magnetic guide field is shown in the right panel of Fig. \ref{fig:line_profile_asym}. In addition to the redshifted major component, there is a blueshifted minor component corresponding to an upflow into the loop associated with a region of increased heating rate.\\
To test whether the enhancement of the line wings increases with the size of the resolution element,
we select a region of $0.5\times 0.5\; \rm{Mm}$ centered around the ray marked in Fig. \ref{fig:apex_cs}. The region is outlined by the black rectangle in Fig. \ref{fig:ntlw_hr}. The line profile averaged over a region of large line-broadening and the underlying velocity distribution for the region along the LOS are shown in Fig. \ref{fig:spec_vel}. 
While the velocity distribution possesses enhancements in the wings, the resulting line profile can be well fitted by a Gaussian. The velocity distribution shown in the right panel shows a slight enhancement in both wings.
The wing excess of the line profile, however, is small at just 2.3 percent.\\
The hottest regions are not necessarily the brightest.
Figure \ref{fig:box_overview_ntlw} shows that whilst the plasma at the location of the heating event contained in the examined region is very hot, it is not the brightest region because the temperature lies above the line formation temperature of \fexv. The line profile is thus dominated by emission from cooler regions with lower velocities. This could explain the lack of enhanced line wings at lower effective resolutions.\\
 
\subsection{Dependence on spatial resolution in synthesized observations}
\label{section:res_dep}

\begin{figure}
\resizebox{\hsize}{!}{\includegraphics{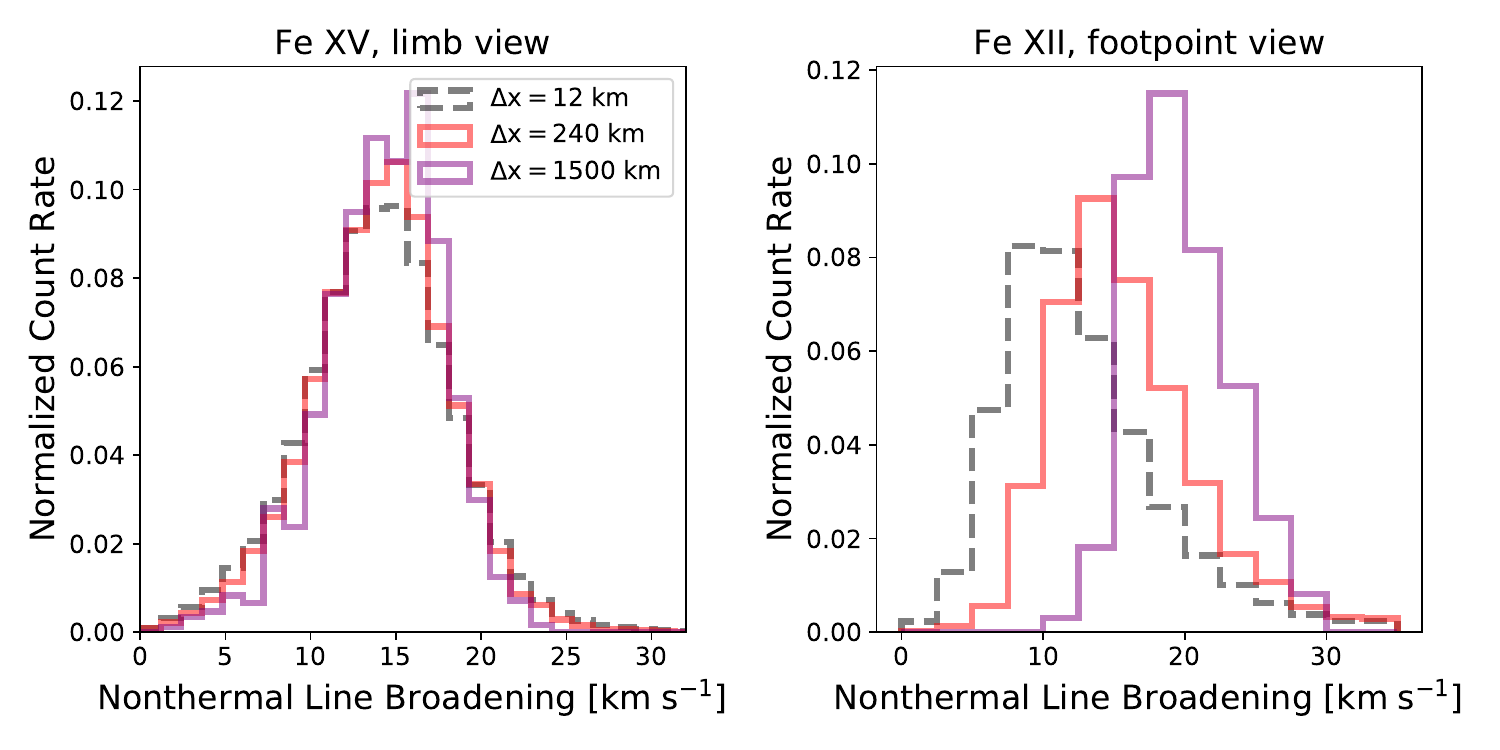}}
  \caption{Distribution of non-thermal line widths for different binning factors and lines of sight. Left panel: Effective line widths for \fexv\ and a LOS perpendicular to the guide field in the x-direction, essentially mimicking a view of the loop at the limb. The behavior for the y-direction is similar. We chose the \fexv\ line for the perpendicular view since a large part of the plasma in the loop is at temperatures in the range of 2-3 MK. Right panel: Effective line width for \fexii\ and a LOS parallel to the guide field for s $\rm{\epsilon}$ [0,10] Mm, mimicking a view of the loop on the disk at its footpoint. For the footpoints we have used the \fexii\ line to account for the cooler plasma there. A lower number of bins is used for the footpoints since the loop cross-section is covered by a smaller number of pixels than the side view. See Sect. \ref{section:res_dep}.}
  \label{fig:histo_rebin}
\end{figure}

We examine both the dependence of the non-thermal line widths on the numerical grid resolution at which the simulation is run as well as on the spatial resolution of the (hypothetically) observing instrument, which is more coarse than the full resolution of the numerical model for existing instruments.
First, the synthetic line profiles are computed from the numerical model at the native resolution of the simulation. The profiles are then rebinned in order to degrade the resolution as described in Sect. \ref{section:non-thermal}. \\
We find higher values for the non-thermal line broadening for a higher numerical grid resolution. The difference between the MR and HR runs is considerably smaller than between the LR and MR runs for both the perpendicular and parallel LOS (see Fig. \ref{fig:hist_gridres} in Appendix \ref{app:gridres}). Observed non-thermal line widths have been found to be independent of the resolution of the observing instrument \citep{2015ApJ...799L..12D}. In order to investigate the dependence on the spatial scales of observations, we apply different levels of spatial binning to the synthesized spectra before calculating the profile moments. For this part of the study, the simulation run with the highest resolution of $\Delta x=12\; \rm{km}$ was employed. The rebinning was performed by summing up the line profiles over a square with a side length of different multiples of the grid spacing.  
In Fig. \ref{fig:histo_rebin}, we show a time-averaged histogram for the non-thermal line widths for different effective resolutions ranging from the original resolution of $\Delta x=12\; \rm{km}$ to $\Delta x=1500\; \rm{km}$. Spatial resolutions for existing spectrometers are $\sim 2^{\prime\prime}$ for the \textit{Hinode}/EIS instrument and $\sim 0.^{\prime\prime}33-0.^{\prime\prime}4$ for IRIS, corresponding to roughly $1450\; \rm{km}$ and $239-290\; \rm{km}$ on the Sun. We have calculated the line widths for five different effective resolutions. For clarity, we include only three histograms. The distribution changes smoothly with decreasing effective resolution.\\
The non-thermal line width is almost independent of resolution for a limb view of the loop. For the line of sight parallel to the magnetic field at the loop footpoints, however, we find that the line broadening increases for a coarser binning, viz. resolution. For the non-thermal line broadening perpendicular to the guide field, the position of the peak of the distribution lies at roughly 15 $\rm{km\; s^{-1}}$, independently of the chosen spatial resolution. For the LOS parallel to the guide field, the peak is shifted towards larger line widths, from about 7.5 $\rm{km\; s^{-1}}$ for the original resolution of 12 $\rm{km}$ to roughly 17.5 $\rm{km\; s^{-1}}$ for the lowest effective resolution of 1500 $\rm{km}$. 
The histograms for the footpoint view have to be treated with caution due the limited width of the box of just six by six megameters. For the lowest resolution considered here, the loop cross section is only covered by 16 pixels for one timestep. \\
A Fourier transform of the different velocity components in a plane at the loop apex
confirms that the transverse components have more power at small scales than the longitudinal component (see Fig. \ref{fig:fourier}). The difference in scale is also seen when comparing $v_{x}$ and $v_{y}$ with $v_{z}$ in Fig. \ref{fig:apex_cs} or Fig. \ref{fig:vel_comp}.\\
This suggests that for the perpendicular LOS, the line broadening is mainly produced by velocity fluctuations along the line of sight, so that line profiles produced by integrating along a single LOS already have the width of profiles averaged over a larger loop area. Length scales for fluctuations in the field-aligned velocity component are larger, thus the effect of including patches of parallel velocities with different magnitudes and opposite sign when decreasing the effective resolution dominates over the broadening from LOS integration alone.\\
We checked the dependence of the non-thermal line width on exposure time for a five minute time sequence of run LR with a cadence of 5 s. We used the \fexv\ emission line and a line of sight perpendicular to the guide field. We used four different exposure times, 300 s, 150 s, 125 s and 15 s, by adding up the snapshots of the model during corresponding time intervals. We did not find a significant dependence of the non-thermal line broadening on the exposure time.
\section{Discussion} 
\label{section:disc_ntlw}

\subsection{Perpendicular broadening}

The relation between the non-thermal broadening and the angle under which the loop structure is observed is important for the determination of the mechanism underlying the broadening of emission lines and the heating of the corona.
For on-disk observations, the middle part of the coronal loop would be mostly observed with a LOS almost perpendicular to the magnetic field, while at the loop footpoints the LOS is more closely aligned with the magnetic guide field.\\
The non-thermal line width reaches values up to and above 30 $\rm{km\; s^{-1}}$ which is consistent with observations \citep{1999ApJ...513..969H,2016ApJ...827...99T,2016ApJ...820...63B}.
For the 1-2 MK corona, \citet{1999ApJ...513..969H} observed non-thermal line widths of 14-20 $\rm{km\; s^{-1}}$ and 10-18 $\rm{km\; s^{-1}}$ in an active region for assumed ion temperatures of 1.0 and 1.8 MK, respectively. The non-thermal line widths observed for very hot plasma at 3.5 MK were 16-24 $\rm{km s^{-1}}$, while \citet{2009ApJ...705L.208I} found non-thermal line widths of 13 $\rm{km\; s^{-1}}$ for active region loop plasma with ion temperatures of 2.5 MK. \citet{2016ApJ...820...63B} estimated the mean non-thermal line widths  to fall between 13.5 and 21.6 $\rm{km s^{-1}}$ for 16 core arcade loops observed on-disk at temperatures ranging from 1-4 MK, with a mean line broadening of 17.6 $\rm{km\; s^{-1}}$. In contrast to this, \citet{2016ApJ...827...99T} measure a non-thermal line width distribution with a peak at $15\; \mathrm{km\; s^{-1}}$ for the \fexii\ 195.119 \AA\ line observed with EIS and the \fexii\ 1349.4 \AA\ line observed by IRIS but higher mean non-thermal line widths of 24 $\rm{km\; s^{-1}}$ for the \fexii\ 1349.4 \AA\ line.\\
The data set of \citet{2016ApJ...827...99T}, however, contains post-flare loops in addition to active region moss, which might lead to the presence of strong mass flows. Therefore caution has to be exercised in comparing the measured line width distribution to non-flaring conditions.\\
The coronal lines used in our study are produced mainly by plasma around 2.2 and 1.5 MK and the line widths we found agree well with the observed widths for a plasma temperature of 1.8 MK by \citet{1999ApJ...513..969H}. 
Overall, the mean non-thermal velocities we find in our study for a LOS perpendicular to the magnetic field are compatible with observed line widths found by \citet{2009ApJ...705L.208I}. For temperatures between 1.8-2.5 MK, \citet{1999ApJ...513..969H} get only slightly higher non-thermal line widths than the values for \fexv\ from our model. Our simulations reproduce the peak at 15 $\rm{km s^{-1}}$ found for the \fexii\ 195.119 \AA\ and 1349.4 \AA\ lines by \citet{2016ApJ...827...99T}, but not the tail at high line widths for the 1349.4 \AA\ line. The most violent events leading to very high line width thus seem to be missing in our simulation.\\
A limitation of our study is the narrow width of the simulation box. We integrate over a range of just six Mm. On the Sun, a loop would never be observed in isolation, more plasma would be present along the LOS in front and behind the loop and potentially contribute to the broadening of the emission lines. 
\\
The non-thermal broadening perpendicular to the magnetic field has been interpreted as resulting from small-scale twist \citep{2015ApJ...799L..12D,2022arXiv220107290R}. Consistently, we find for run HR that there is a moderate correlation between the non-thermal line broadening for \fexv\ and the LOS-averaged unsigned axial vorticity.

\subsection{Parallel broadening}

In reality, coronal loops are observed from a variety of angles between the LOS and the loop axis. The LOS is oriented predominantly perpendicular to the loop axis for loops seen face-on at the limb and at the loop apex for on-disk observations.\\
The LOS is parallel to the loop axis for observations of the loop footpoints for on-disk observations and near the apex for loops observed edge-on at the limb.\\
To isolate the footpoint region, we have integrated the spectra for the footpoints up to a distance along the loop of s=10 Mm. 
While of the same order of magnitude, peak and mean non-thermal line widths are smaller for a LOS parallel to the guide field at the loop footpoints than for a LOS perpendicular to the guide field in our study. While the distribution of non-thermal line widths seen perpendicular to the guide field is nearly symmetric, the distribution for the non-thermal line widths seen parallel to the loop axis show a high-velocity tail. 
\\
The non-thermal velocities seen along the guide field at the footpoints are larger than those seen in previous straightened loop models. 
In the study by \citet{2020A&A...639A..21P}, the perpendicular velocity fluctuations are significantly larger than the fluctuations aligned with the field. 
The model of \citet{2020A&A...639A..21P} does not include a chromosphere that could serve as mass reservoir for upflows induced by coronal heating. The non-thermal line widths in the parallel direction are thus significantly smaller than in our model.\\
Coupling a straightened loop simulation to a realistic lower atmosphere, as we do here, allows for chromospheric evaporation in response to heating events and thus leads to larger field-aligned velocities.\\
The connection between chromospheric evaporation and non-thermal line width is illustrated in
Fig. \ref{fig:time_evo_cube}. The vertical velocity starts to increase with a delay after the rise in the heating rate. A similar behavior is seen for the mass flux shown in Fig. \ref{fig:time_evo_par}. The mass flux in the low corona reaches a peak approximately one minute after the occurrence of the strongest heating event. This points to chromospheric evaporation in response to heating events as a cause for strong field-aligned upflows. The influx of mass into the loop is present at both footpoints. Heat is conducted along the magnetic field away from the heating event down to both footpoints, therefore finding an increased mass flux at each footpoint is expected for chromospheric evaporation. The mass flux at each loop footpoint peaks roughly 30 s apart. The speed of a propagating heat front can be the fastest speed in the simulation. MURaM makes use of a hyperbolic formulation of the heat equation in order to avoid very small timesteps. As a result, the propagating heat front is artificially slowed down to a speed comparable to the maximum wave speed in the simulation \citep{2017ApJ...834...10R}. The maximum wave speed in the LR run is $\approx 3600\; \rm{km\; s^{-1}}$, leading to a travel time between the two slices under consideration (5 Mm and 45 Mm) of 11 s. The heat front therefore reaches the dense chromosphere at each loop footpoint at slightly different times. Since the heating event is closer to the first footpoint, this delay could partially explain the time difference in the mass flux peaks.

\subsection{Perpendicular and parallel broadening}

 Different coronal heating theories lead to distinct predictions for the non-thermal line width as a function of the angle between the LOS and the loop axis.
While Alfv\'{e}n wave models predict larger non-thermal line widths for a loop seen perpendicular to the loop axis due to energy transport by perturbations transverse to the guide field, magnetoacoustic waves would lead to larger broadening parallel to the loop axis. Nanoflares are thought to lead to strong flows due to evaporation of plasma from the chromosphere and thus higher line widths parallel to the magnetic field.
The different origins of the line broadening parallel and perpendicular to the magnetic field should lead to a center-to-limb variation of the non-thermal line broadening.\\
Most existing numerical studies have not explicitly examined the dependence of the non-thermal line width on the observing angle.
Results from observational studies do not always give mutually consistent results on the center-to-limb variation of nonthermal line widths and the variation along individual structures. \citet{2016ApJ...827...99T} find the largest non-thermal velocities with EIS at the footpoints of large fan loops, while moss regions show smaller non-thermal motion. Regarding the center-to-limb variation, they find more regions with high nonthermal velocities for on-disk observations and conclude that field-aligned flows play an important role for the broadening of emission lines. \citet{2008ApJ...678L..67H} also measure a decrease of nonthermal line width towards the loop top near the limb. Other studies find higher non-thermal velocities at the limb (\citet{1998ApJ...505..957C,1998A&A...337..287E,2022arXiv220107290R}), the former two finding only very small changes for coronal lines.
\\
\citet{1999ApJ...513..969H} conducted a study on the variation of the non-thermal line width along individual loops seen from different angles, comparing loops observed perpendicular and parallel to the loop axis at the limb. They found a slight decrease of the non-thermal line width towards the apex for the edge-on loops with a viewing direction parallel to the magnetic field at the top for some coronal lines. For the loops seen perpendicular to the loop axis, they find an increase towards the loop top. This behavior suggests that flows perpendicular to the magnetic field contribute more to the line broadening than field-aligned flows. If the line profile is fit with multiple Gaussians, the line width increases towards the loop top if the individual components are taken into account \citep{2010A&A...521A..51P}. In this study, we did not decompose the spectral profiles into several components, so that a direct comparison with the results of \citet{2010A&A...521A..51P} is not possible.\\
We focus here mainly on the non-thermal line widths seen perpendicular to the loop axis and at the loop footpoints, since we chose the angles under which different loop parts are most likely to be seen.\\
While the non-thermal line widths at the footpoints are slightly smaller, the velocity distributions for the velocity components perpendicular and parallel to the guide field are anisotropic, with the parallel velocity component reaching larger magnitudes. This can be seen in Fig. \ref{fig:spec_vel} for a $0.5\times 0.5 \times 6$ Mm box centered on a region with large line broadening at the apex. The lower line broadening at the footpoints seen parallel to the loop axis are most likely due to the bulk part of the emission stemming from denser and slower moving plasma as well as the larger length scales of structures along the loop axis. High plasma velocities therefore lead to high Doppler shifts, but not necessarily to high non-thermal line widths. For the LOS perpendicular to the magnetic field, we find a more symmetrical distribution.\\
In order to reproduce observed non-thermal line broadening, simulations need to capture both small-scale motions perpendicular to the guide field and chromospheric evaporation along the field lines.
We find larger simulated non-thermal line widths than \citet{2016ApJ...827...99T} computed for a Bifrost snapshot, however, we find comparable values to \citet{2015ApJ...802....5O} for the \fexii\ line. \citet{2010ApJ...718.1070H} generally found the highest non-thermal line widths near the footpoints of their simulated loop arcade. In contrast to this, \citet{2015ApJ...802....5O} found smaller values for the line widths at the footpoints for coronal lines than for the apex or the line profiles averaged over the entire simulation domain. 
In the simulation snapshot used by \citet{2016ApJ...827...99T}, coronal temperatures reach very high values, the \fexii\ emission is therefore concentrated closer to the footpoints, which could explain the lower velocities. 
A detailed study of the variation of the non-thermal line width along the loop axis is the subject of future work. \\
The loop-averaged line widths for run LR for a LOS parallel to the guide field reach larger values than seen in the perpendicular direction (See Fig. \ref{fig:time_evo_par}). This is in contrast to the statistical distributions of the non-thermal line widths shown in Fig. \ref{fig:hist_Fe15_LOS}, which yield smaller average non-thermal line widths parallel to the loop axis at the original spatial resolution. This can be explained by the different relation between intensity and non-thermal line width for the different LOS. The line formation temperature for \fexv\ is roughly 2 MK, but the hottest plasma in the regions with the strongest heating and highest velocities reaches temperatures of more than four MK. Thus, the hottest regions with the largest line broadening are not the brightest regions. When computing the loop-averaged line broadening by summing up the line profiles and extracting the non-thermal line widths, the cooler regions of the loop are weighted more strongly. \\
Combining the effects of the parallel line broadening (near the footprints) and the perpendicular broadening (near the apex) we can make some prediction on how the non-thermal broadening should change along the loop from footprint to apex. When observing a loop on the disk, one would expect that the broadening is increasing from the footprint to the apex. This would be consistent with the observational study of \citet{2010A&A...521A..51P} when considering only the broadening of the core component of the line profile. However further careful observational studies would be needed to draw a final conclusion on this.\\
In order to study the effect of the observing angle on the measured non-thermal line widths in observations, ideally the same loop system needs to be observed from different vantage points. The SPICE instrument on Solar Orbiter \citep{2020A&A...642A..14S} in addition to EIS would present an option, but unfortunately SPICE does not have sufficient spectral resolution to accurately determine the line widths in non-flaring coronal loops.\\\\
A possible way to study the variation of non-thermal line widths without the availability of two-point observations is a statistical study of the center-to-limb variation of the non-thermal line widths in a large number of ARs analogous to the study conducted in \citet{2022A&A...660A...3M} for transition region lines. \citet{2022A&A...660A...3M} find an anisotropy for the non-thermal line widths for the transition region in ARs, with a larger vertical than horizontal component.
This is in contrast to the line width distribution for the loop footpoints in our simulation. However, it is unclear how their results for the cool transition region emission originating from $10^5$ K in probably low-lying structures relates to to the coronal loops we investigate here. Similar studies with a large number of active regions have not been performed to the same extent for coronal lines. Near-limb non-thermal velocities found by \citet{1998ApJ...505..957C} exceed disk center values by 2-3 $\rm{km\; s^{-1}}$ for all studied lines apart from O I and Mg X. This is compatible with the difference we find between the average values for a line-of-sight perpendicular to the loop guide field and parallel to the loop guide field. We also find that the nonthermal velocities increase with decreasing spatial instrumental resolution for the parallel line of sight, so we expect even smaller differences at the resolution of SUMER (1000 km). Note that both the on-disk as well as the near-limb spectra contain contributions from flows along and perpendicular to the magnetic field.
While \citet{1998ApJ...505..957C} do not study variations of nonthermal velocities along individual structures,
\citet{1999ApJ...513..969H} find a decrease of nonthermal velocities near the loop top by 3-5 $\rm{km\; s^{-1}}$ for the green coronal line if the viewing direction is nearly parallel to the magnetic field at the top. This finding is compatible with the higher nonthermal velocities we find for the perpendicular viewing direction.
\\
A possible explanation for our synthesized non-thermal line broadening values not reaching the highest observed values is our setup with a strong uniform guide field, corresponding to a long loop with very low-lying, small-scale loops near the footpoints. In our simulation, the non-thermal broadening stems from small-scale velocity fluctuations along the LOS. Effects such as large-scale flux emergence or interaction of several loop systems are not taken into account. These events might lead to a stronger acceleration of plasma.  \\
Overall, while previous studies using the same loop geometry focused on specific effects such as line broadening from unwinding of a magnetic braid \citep{2020A&A...639A..21P} or injected waves \citep{2019ApJ...881...95P,2014ApJ...786...28A}, our simulation combines a realistic driver without the need for artificial forcing, flows along the magnetic field due to the inclusion of a chromosphere, and high resolution across the loop to capture broadening by small-scale motions. By this, we get quite close to the observed values for the observed non-thermal broadening, based on a self-consistent driving of the coronal loop.
\subsection{Line profiles}

Line profiles with excess emission in the wings have been found for example in \citet{2010A&A...521A..51P}.
In \citet{2020A&A...639A..21P} enhanced wings in the velocity distribution are interpreted as a sign of turbulence. While we find complex motions on various scales in the loop interior, the motions are less violent than in the case of \citet{2020A&A...639A..21P} where the untwisting braid leads to large velocities. Instead of an unwinding of a pre-braided field, the magnetic field in our simulation is continuously driven by magnetoconvection. We therefore do not have large-scale unbraiding events driving strong flows. The small wing excess is similar to the braid relaxation simulation at later times in the simulation when the initial turbulent state has already partially decayed.

\subsection{Heating}
The averaged heating rate shows intermittent strong heating events.
The increase in density in response to the elevated temperature is delayed because material needs time to reach the corona through chromospheric evaporation.\\
The non-thermal line broadening is often used as a proxy for the heating of the corona.
In the timeseries, the non-thermal velocities seen perpendicular to the guide field and partially the non-thermal velocities parallel to the guide field follow the time evolution of the heating rate. \\
Similar to \citet{2020A&A...639A..21P}, we find that peaks in the line width are often preceded by a drop in the magnetic energy associated with the transverse field components. Unwinding and reconnection of the magnetic field could thus cause the first two heating events shown in Fig. \ref{fig:time_evo} at 17.9 min and 30 min. The excess magnetic energy, however, is replenished by shuffling of the magnetic field due to magnetoconvection. The line width is thus always significantly above the thermal line width.\\
At the time of largest broadening in the low resolution timeseries, we find several reconnection events heating the plasma and leading to outflows. We can directly attribute the largest peak in heating and non-thermal line width to a reconnection event accelerating plasma, which is shown in Fig. \ref{fig:vel_comp}. The heating event occurs at the location of shearing of several misaligned bundles of magnetic field lines (see Fig. \ref{fig:braid}). In addition to an enhanced transverse velocity, a strong upflow is present, indicating that the heat deposited at the reconnection site drives chromospheric evaporation that leads to increased line broadening parallel to the magnetic field.\\
A similar event occurs in run HR at x,y=[1.2,4] Mm,
leading to an asymmetric line profile (see Fig. \ref{fig:line_profile_asym}).
This event is reminiscent of the reconnection nanojets in \citet{2021NatAs...5...54A}. The outflow associated with this reconnection extends over a vertical range of 4 Mm, but is strongly collimated in the direction transverse to the guide field. In our simulation, the outflow has a width of 80 km at the apex.
The velocities in the bidirectional outflow at the apex go up to 155 $\rm{km\; s^{-1}}$. 
For the examined heating event in the low resolution time series, we find that a rise in the transverse velocity precedes the sharp increase in the heating rate. The increase in transverse velocity could thus be associated with both an increased Poynting flux that is partially dissipated in the heating event, or be a response to the heating event due to outflows from the reconnection site.\\
For the high resolution case, we find a weak to moderate correlation between non-thermal broadening and both Poynting flux, vorticity and heating rate. non-thermal line broadening has been interpreted as an indication for both Alfv\'{e}n waves as well as flows developing in response to nanoflare heating \citep{2006ApJ...647.1452P}. While we find bursty heating events associated with reconnection after the buildup of magnetic energy over timescales of the order of ten minutes, this does not exclude that waves are also present.
\subsection{Dependence on resolution}

While the HR run with a grid spacing of 12 km seems to have a good resolution to study non-thermal line broadening, even at the highest numerical resolution that we employ the non-thermal line width does not seem to have completely converged. We therefore expect slightly higher line widths for a higher grid resolution, bringing the results closer to observations.\\
The behavior of the broadening in the direction perpendicular to the guide field for different effective instrument resolutions is consistent with the invariance to spatial rebinning found by \citet{2015ApJ...799L..12D}.
The maximal rebinning factor used in \citet{2015ApJ...799L..12D} and \citet{2016ApJ...827...99T} is 12, corresponding to a spatial resolution of $1520\times 250.8\; \rm{km}$ while we examine in our study a change of resolution by a factor of $\sim 100$ and find that the invariance still holds.
\citet{2015ApJ...802....5O} found that the non-thermal line width increases by about 20 \% with spatial rebinning to a resolution of $1^{\prime\prime}$, however, the magnetic field in their simulation box has a complex shape and therefore both field-aligned and perpendicular motions are present along the LOS if the loop system is seen from above.
The independence of the line broadening perpendicular to the guide field and the slight dependence of the line broadening parallel to the guide field can be explained by the different sizes of structures in the transverse and field aligned velocity components. 
As shown in the cut through the loop cross-section in Fig. \ref{fig:apex_cs}, the transverse velocity components have a complex small-scale structure while structures in the field-aligned velocity are larger. A Fourier transform of the unsigned velocity components at the loop apex confirms that the field-aligned velocity has more power on larger scales above roughly 200 km and falls off faster than the power spectra for the transverse velocity components below this value.\\
For sufficiently fine effective resolutions, the resolution element is still smaller than typical structure in the field-aligned velocity. For pixels larger than 240 km, one pixel in the synthetic observation covers several structures with different velocities along the LOS.
For the LOS perpendicular to the guide field, the LOS-integration over the width of the simulation box of six Mm leads to broadening due to small-scale velocity components along the LOS even for a resolution element comprised of just one grid cell with a side length of 12 km. For the parallel velocity component, structures are elongated along the line of sight and show less fluctuations. Integration over a field of view larger than the grid cells thus leads to increased broadening.
Even with significantly increased resolution, instruments would likely not see a large difference in the non-thermal line broadening when observing across the loop due to the LOS-integration. The dependence of the parallel broadening on the FOV is difficult to test since normally the magnetic field would not align perfectly with the line of sight. This effect could possibly be observed in moss regions where the field is mostly vertical.
\section{Conclusion}
\label{section:conc}

In our numerical model, we find values for the non-thermal line broadening both perpendicular and parallel to the guide field that are compatible with observations.\\
We include proper treatment of self-consistent magneto-convection in the photosphere and of the mass transfer between the chromosphere and corona. The treatment of the coupling between the different atmospheric layers is necessary to properly account for field-aligned flows. Thus we arrive at realistic values for the non-thermal line broadening in the field-parallel direction along the loop axis if the limited resolution of observing instruments is taken into account. Including a shallow convection zone also ensures that the non-thermal line broadening is sustained throughout the lifetime of the loop by continuous driving.
The line broadening measured perpendicular and parallel to the loop axis is of similar order of magnitude, but arises from different causes. For a LOS perpendicular to the loop axis, the spectral line profiles are broadened due to the formation of small-scale MHD turbulent flows perpendicular to the guide field, including outflows from heating events. For a LOS parallel to the guide field, we find the largest values for the non-thermal broadening due to evaporative upflows developing in response to heating events.
The non-thermal line broadening follows the time evolution of the heating rate. For a single snapshot in time, there is a weak to moderate correlation between heating rate and non-thermal line broadening. We can directly relate the largest values of non-thermal line broadening to strong heating events and resulting outflows.\\
Consistent with observations, the non-thermal line broadening is independent of effective resolution for the line of sight perpendicular to the guide field. For the LOS parallel to the guide field, the non-thermal line broadening increases with decreasing effective resolution. In conclusion, our model of a stretched loop based on a self-consistent 3D MHD model from the upper convection zone into the corona provides a comprehensive explanation of the non-thermal broadening observed in coronal loops.

\section{Acknowledgements}

This project has received funding from the European
Research Council (ERC) under the European Union’s Horizon 2020 research
and innovation program (grant agreement No. 695075). We gratefully acknowledge the computational resources provided by the Cobra supercomputer system of the Max Planck Computing and Data Facility (MPCDF) in Garching,
Germany. 
The research leading to these results has received funding from the UK Science and Technology Facilities Council (consolidated grant ST/W001195/1).
Ineke De Moortel received funding from the Research Council of Norway through its Centres of Excellence scheme, project number 262622.
CHIANTI is a collaborative project involving George Mason University, the University of Michigan (USA), University of Cambridge (UK) and NASA Goddard Space Flight Center (USA).

\section*{Data Availability}

The data from the numerical simulations and analysis presented
in this paper are available from the corresponding author upon
reasonable request.

\bibliographystyle{mnras}
\bibliography{paper}

\begin{thebibliography}{}
\makeatletter
\relax
\def\mn@urlcharsother{\let\do\@makeother \do\$\do\&\do\#\do\^\do\_\do\%\do\~}
\def\mn@doi{\begingroup\mn@urlcharsother \@ifnextchar [ {\mn@doi@} {\mn@doi@[]}}
\def\mn@doi@[#1]#2{\def\@tempa{#1}\ifx\@tempa\@empty \href {http://dx.doi.org/#2} {doi:#2}\else \href {http://dx.doi.org/#2} {#1}\fi \endgroup}
\def\mn@eprint#1#2{\mn@eprint@#1:#2::\@nil}
\def\mn@eprint@arXiv#1{\href {http://arxiv.org/abs/#1} {{\tt arXiv:#1}}}
\def\mn@eprint@dblp#1{\href {http://dblp.uni-trier.de/rec/bibtex/#1.xml} {dblp:#1}}
\def\mn@eprint@#1:#2:#3:#4\@nil{\def\@tempa {#1}\def\@tempb {#2}\def\@tempc {#3}\ifx \@tempc \@empty \let \@tempc \@tempb \let \@tempb \@tempa \fi \ifx \@tempb \@empty \def\@tempb {arXiv}\fi \@ifundefined {mn@eprint@\@tempb}{\@tempb:\@tempc}{\expandafter \expandafter \csname mn@eprint@\@tempb\endcsname \expandafter{\@tempc}}}

\bibitem[\protect\citeauthoryear{{Antolin}, {Pagano}, {Testa}, {Petralia}  \& {Reale}}{{Antolin} et~al.}{2021}]{2021NatAs...5...54A}
{Antolin} P.,  {Pagano} P.,  {Testa} P.,  {Petralia} A.,   {Reale} F.,  2021, \mn@doi [Nature Astronomy] {10.1038/s41550-020-1199-8}, \href {https://ui.adsabs.harvard.edu/abs/2021NatAs...5...54A} {5, 54}

\bibitem[\protect\citeauthoryear{{Asgari-Targhi}, {van Ballegooijen}  \& {Imada}}{{Asgari-Targhi} et~al.}{2014}]{2014ApJ...786...28A}
{Asgari-Targhi} M.,  {van Ballegooijen} A.~A.,   {Imada} S.,  2014, \mn@doi [\apj] {10.1088/0004-637X/786/1/28}, \href {https://ui.adsabs.harvard.edu/abs/2014ApJ...786...28A} {786, 28}

\bibitem[\protect\citeauthoryear{{Breu}, {Peter}, {Cameron}, {Solanki}, {Przybylski}, {Rempel}  \& {Chitta}}{{Breu} et~al.}{2022}]{2022A&A...658A..45B}
{Breu} C.,  {Peter} H.,  {Cameron} R.,  {Solanki} S.~K.,  {Przybylski} D.,  {Rempel} M.,   {Chitta} L.~P.,  2022, \mn@doi [\aap] {10.1051/0004-6361/202141451}, \href {https://ui.adsabs.harvard.edu/abs/2022A&A...658A..45B} {658, A45}

\bibitem[\protect\citeauthoryear{{Brooks} \& {Warren}}{{Brooks} \& {Warren}}{2016}]{2016ApJ...820...63B}
{Brooks} D.~H.,  {Warren} H.~P.,  2016, \mn@doi [\apj] {10.3847/0004-637X/820/1/63}, \href {https://ui.adsabs.harvard.edu/abs/2016ApJ...820...63B} {820, 63}

\bibitem[\protect\citeauthoryear{{Chae}, {Sch{\"u}hle}  \& {Lemaire}}{{Chae} et~al.}{1998}]{1998ApJ...505..957C}
{Chae} J.,  {Sch{\"u}hle} U.,   {Lemaire} P.,  1998, \mn@doi [\apj] {10.1086/306179}, \href {https://ui.adsabs.harvard.edu/abs/1998ApJ...505..957C} {505, 957}

\bibitem[\protect\citeauthoryear{{Culhane} et~al.,}{{Culhane} et~al.}{2007}]{2007SoPh..243...19C}
{Culhane} J.~L.,  et~al., 2007, \mn@doi [\solphys] {10.1007/s01007-007-0293-1}, \href {https://ui.adsabs.harvard.edu/abs/2007SoPh..243...19C} {243, 19}

\bibitem[\protect\citeauthoryear{{De Pontieu} et~al.,}{{De Pontieu} et~al.}{2014}]{2014SoPh..289.2733D}
{De Pontieu} B.,  et~al., 2014, \mn@doi [\solphys] {10.1007/s11207-014-0485-y}, \href {https://ui.adsabs.harvard.edu/abs/2014SoPh..289.2733D} {289, 2733}

\bibitem[\protect\citeauthoryear{{De Pontieu}, {McIntosh}, {Martinez-Sykora}, {Peter}  \& {Pereira}}{{De Pontieu} et~al.}{2015}]{2015ApJ...799L..12D}
{De Pontieu} B.,  {McIntosh} S.,  {Martinez-Sykora} J.,  {Peter} H.,   {Pereira} T.~M.~D.,  2015, \mn@doi [\apjl] {10.1088/2041-8205/799/1/L12}, \href {https://ui.adsabs.harvard.edu/abs/2015ApJ...799L..12D} {799, L12}

\bibitem[\protect\citeauthoryear{{Del Zanna}, {Dere}, {Young}  \& {Landi}}{{Del Zanna} et~al.}{2021}]{2021ApJ...909...38D}
{Del Zanna} G.,  {Dere} K.~P.,  {Young} P.~R.,   {Landi} E.,  2021, \mn@doi [\apj] {10.3847/1538-4357/abd8ce}, \href {https://ui.adsabs.harvard.edu/abs/2021ApJ...909...38D} {909, 38}

\bibitem[\protect\citeauthoryear{{Dere} \& {Mason}}{{Dere} \& {Mason}}{1993}]{1993SoPh..144..217D}
{Dere} K.~P.,  {Mason} H.~E.,  1993, \mn@doi [\solphys] {10.1007/BF00627590}, \href {https://ui.adsabs.harvard.edu/abs/1993SoPh..144..217D} {144, 217}

\bibitem[\protect\citeauthoryear{{Dere}, {Bartoe}  \& {Brueckner}}{{Dere} et~al.}{1984}]{1984ApJ...281..870D}
{Dere} K.~P.,  {Bartoe} J. D.~F.,   {Brueckner} G.~E.,  1984, \mn@doi [\apj] {10.1086/162167}, \href {https://ui.adsabs.harvard.edu/abs/1984ApJ...281..870D} {281, 870}

\bibitem[\protect\citeauthoryear{{Dere}, {Landi}, {Mason}, {Monsignori Fossi}  \& {Young}}{{Dere} et~al.}{1997}]{1997A&AS..125..149D}
{Dere} K.~P.,  {Landi} E.,  {Mason} H.~E.,  {Monsignori Fossi} B.~C.,   {Young} P.~R.,  1997, \mn@doi [\aaps] {10.1051/aas:1997368}, \href {https://ui.adsabs.harvard.edu/abs/1997A&AS..125..149D} {125, 149}

\bibitem[\protect\citeauthoryear{{Erdelyi}, {Doyle}, {Perez}  \& {Wilhelm}}{{Erdelyi} et~al.}{1998}]{1998A&A...337..287E}
{Erdelyi} R.,  {Doyle} J.~G.,  {Perez} M.~E.,   {Wilhelm} K.,  1998, \aap, \href {https://ui.adsabs.harvard.edu/abs/1998A&A...337..287E} {337, 287}

\bibitem[\protect\citeauthoryear{{Gordovskyy}, {Kontar}  \& {Browning}}{{Gordovskyy} et~al.}{2016}]{2016A&A...589A.104G}
{Gordovskyy} M.,  {Kontar} E.~P.,   {Browning} P.~K.,  2016, \mn@doi [\aap] {10.1051/0004-6361/201527249}, \href {https://ui.adsabs.harvard.edu/abs/2016A&A...589A.104G} {589, A104}

\bibitem[\protect\citeauthoryear{{Hansteen}, {Hara}, {De Pontieu}  \& {Carlsson}}{{Hansteen} et~al.}{2010}]{2010ApJ...718.1070H}
{Hansteen} V.~H.,  {Hara} H.,  {De Pontieu} B.,   {Carlsson} M.,  2010, \mn@doi [\apj] {10.1088/0004-637X/718/2/1070}, \href {https://ui.adsabs.harvard.edu/abs/2010ApJ...718.1070H} {718, 1070}

\bibitem[\protect\citeauthoryear{{Hara} \& {Ichimoto}}{{Hara} \& {Ichimoto}}{1999}]{1999ApJ...513..969H}
{Hara} H.,  {Ichimoto} K.,  1999, \mn@doi [\apj] {10.1086/306880}, \href {https://ui.adsabs.harvard.edu/abs/1999ApJ...513..969H} {513, 969}

\bibitem[\protect\citeauthoryear{{Hara}, {Watanabe}, {Harra}, {Culhane}, {Young}, {Mariska}  \& {Doschek}}{{Hara} et~al.}{2008}]{2008ApJ...678L..67H}
{Hara} H.,  {Watanabe} T.,  {Harra} L.~K.,  {Culhane} J.~L.,  {Young} P.~R.,  {Mariska} J.~T.,   {Doschek} G.~A.,  2008, \mn@doi [\apjl] {10.1086/588252}, \href {https://ui.adsabs.harvard.edu/abs/2008ApJ...678L..67H} {678, L67}

\bibitem[\protect\citeauthoryear{{Imada}, {Hara}  \& {Watanabe}}{{Imada} et~al.}{2009}]{2009ApJ...705L.208I}
{Imada} S.,  {Hara} H.,   {Watanabe} T.,  2009, \mn@doi [\apjl] {10.1088/0004-637X/705/2/L208}, \href {https://ui.adsabs.harvard.edu/abs/2009ApJ...705L.208I} {705, L208}

\bibitem[\protect\citeauthoryear{{Innes}, {Inhester}, {Axford}  \& {Wilhelm}}{{Innes} et~al.}{1997}]{1997Natur.386..811I}
{Innes} D.~E.,  {Inhester} B.,  {Axford} W.~I.,   {Wilhelm} K.,  1997, \mn@doi [\nat] {10.1038/386811a0}, \href {https://ui.adsabs.harvard.edu/abs/1997Natur.386..811I} {386, 811}

\bibitem[\protect\citeauthoryear{{Kjeldseth Moe} \& {Nicolas}}{{Kjeldseth Moe} \& {Nicolas}}{1977}]{1977ApJ...211..579K}
{Kjeldseth Moe} O.,  {Nicolas} K.~R.,  1977, \mn@doi [\apj] {10.1086/154966}, \href {https://ui.adsabs.harvard.edu/abs/1977ApJ...211..579K} {211, 579}

\bibitem[\protect\citeauthoryear{{Li} \& {Peter}}{{Li} \& {Peter}}{2019}]{2019A&A...626A..98L}
{Li} L.~P.,  {Peter} H.,  2019, \mn@doi [\aap] {10.1051/0004-6361/201935165}, \href {https://ui.adsabs.harvard.edu/abs/2019A&A...626A..98L} {626, A98}

\bibitem[\protect\citeauthoryear{{Matsumoto}}{{Matsumoto}}{2018}]{2018MNRAS.476.3328M}
{Matsumoto} T.,  2018, \mn@doi [\mnras] {10.1093/mnras/sty490}, \href {https://ui.adsabs.harvard.edu/abs/2018MNRAS.476.3328M} {476, 3328}

\bibitem[\protect\citeauthoryear{{McIntosh} \& {De Pontieu}}{{McIntosh} \& {De Pontieu}}{2012}]{2012ApJ...761..138M}
{McIntosh} S.~W.,  {De Pontieu} B.,  2012, \mn@doi [\apj] {10.1088/0004-637X/761/2/138}, \href {https://ui.adsabs.harvard.edu/abs/2012ApJ...761..138M} {761, 138}

\bibitem[\protect\citeauthoryear{{Mou}, {Peter}, {Xia}  \& {Huang}}{{Mou} et~al.}{2022}]{2022A&A...660A...3M}
{Mou} C.,  {Peter} H.,  {Xia} L.,   {Huang} Z.,  2022, \mn@doi [\aap] {10.1051/0004-6361/202142285}, \href {https://ui.adsabs.harvard.edu/abs/2022A&A...660A...3M} {660, A3}

\bibitem[\protect\citeauthoryear{{Olluri}, {Gudiksen}, {Hansteen}  \& {De Pontieu}}{{Olluri} et~al.}{2015}]{2015ApJ...802....5O}
{Olluri} K.,  {Gudiksen} B.~V.,  {Hansteen} V.~H.,   {De Pontieu} B.,  2015, \mn@doi [\apj] {10.1088/0004-637X/802/1/5}, \href {https://ui.adsabs.harvard.edu/abs/2015ApJ...802....5O} {802, 5}

\bibitem[\protect\citeauthoryear{{Pant}, {Magyar}, {Van Doorsselaere}  \& {Morton}}{{Pant} et~al.}{2019}]{2019ApJ...881...95P}
{Pant} V.,  {Magyar} N.,  {Van Doorsselaere} T.,   {Morton} R.~J.,  2019, \mn@doi [\apj] {10.3847/1538-4357/ab2da3}, \href {https://ui.adsabs.harvard.edu/abs/2019ApJ...881...95P} {881, 95}

\bibitem[\protect\citeauthoryear{{Parker}}{{Parker}}{1988}]{1988ApJ...330..474P}
{Parker} E.~N.,  1988, \mn@doi [\apj] {10.1086/166485}, \href {https://ui.adsabs.harvard.edu/abs/1988ApJ...330..474P} {330, 474}

\bibitem[\protect\citeauthoryear{{Patsourakos} \& {Klimchuk}}{{Patsourakos} \& {Klimchuk}}{2006}]{2006ApJ...647.1452P}
{Patsourakos} S.,  {Klimchuk} J.~A.,  2006, \mn@doi [\apj] {10.1086/505517}, \href {https://ui.adsabs.harvard.edu/abs/2006ApJ...647.1452P} {647, 1452}

\bibitem[\protect\citeauthoryear{{Peter}}{{Peter}}{2000}]{2000A&A...360..761P}
{Peter} H.,  2000, \aap, \href {https://ui.adsabs.harvard.edu/abs/2000A&A...360..761P} {360, 761}

\bibitem[\protect\citeauthoryear{{Peter}}{{Peter}}{2010}]{2010A&A...521A..51P}
{Peter} H.,  2010, \mn@doi [\aap] {10.1051/0004-6361/201014433}, \href {https://ui.adsabs.harvard.edu/abs/2010A&A...521A..51P} {521, A51}

\bibitem[\protect\citeauthoryear{{Peter}, {Gudiksen}  \& {Nordlund}}{{Peter} et~al.}{2006}]{2006ApJ...638.1086P}
{Peter} H.,  {Gudiksen} B.~V.,   {Nordlund} {\r{A}}.,  2006, \mn@doi [\apj] {10.1086/499117}, \href {https://ui.adsabs.harvard.edu/abs/2006ApJ...638.1086P} {638, 1086}

\bibitem[\protect\citeauthoryear{{Pontin}, {Peter}  \& {Chitta}}{{Pontin} et~al.}{2020}]{2020A&A...639A..21P}
{Pontin} D.~I.,  {Peter} H.,   {Chitta} L.~P.,  2020, \mn@doi [\aap] {10.1051/0004-6361/202037582}, \href {https://ui.adsabs.harvard.edu/abs/2020A&A...639A..21P} {639, A21}

\bibitem[\protect\citeauthoryear{{Rao}, {Del Zanna}  \& {Mason}}{{Rao} et~al.}{2022}]{2022arXiv220107290R}
{Rao} Y.~K.,  {Del Zanna} G.,   {Mason} H.~E.,  2022, arXiv e-prints, \href {https://ui.adsabs.harvard.edu/abs/2022arXiv220107290R} {p. arXiv:2201.07290}

\bibitem[\protect\citeauthoryear{{Rempel}}{{Rempel}}{2017}]{2017ApJ...834...10R}
{Rempel} M.,  2017, \mn@doi [\apj] {10.3847/1538-4357/834/1/10}, \href {https://ui.adsabs.harvard.edu/abs/2017ApJ...834...10R} {834, 10}

\bibitem[\protect\citeauthoryear{{SPICE Consortium} et~al.,}{{SPICE Consortium} et~al.}{2020}]{2020A&A...642A..14S}
{SPICE Consortium} et~al., 2020, \mn@doi [\aap] {10.1051/0004-6361/201935574}, \href {https://ui.adsabs.harvard.edu/abs/2020A&A...642A..14S} {642, A14}

\bibitem[\protect\citeauthoryear{{Testa}, {De Pontieu}  \& {Hansteen}}{{Testa} et~al.}{2016}]{2016ApJ...827...99T}
{Testa} P.,  {De Pontieu} B.,   {Hansteen} V.,  2016, \mn@doi [\apj] {10.3847/0004-637X/827/2/99}, \href {https://ui.adsabs.harvard.edu/abs/2016ApJ...827...99T} {827, 99}

\bibitem[\protect\citeauthoryear{{V{\"o}gler}}{{V{\"o}gler}}{2003}]{2003PhDT........61V}
{V{\"o}gler} A.,  2003, PhD thesis, Georg August University of Gottingen, Germany

\bibitem[\protect\citeauthoryear{{V{\"o}gler}, {Shelyag}, {Sch{\"u}ssler}, {Cattaneo}, {Emonet}  \& {Linde}}{{V{\"o}gler} et~al.}{2005}]{2005A&A...429..335V}
{V{\"o}gler} A.,  {Shelyag} S.,  {Sch{\"u}ssler} M.,  {Cattaneo} F.,  {Emonet} T.,   {Linde} T.,  2005, \mn@doi [\aap] {10.1051/0004-6361:20041507}, \href {https://ui.adsabs.harvard.edu/abs/2005A&A...429..335V} {429, 335}

\makeatother
\end{thebibliography}

\begin{appendix}

\section{Dependence of the line broadening on grid resolution}
\label{app:gridres}

\begin{figure}
\resizebox{\hsize}{!}{\includegraphics{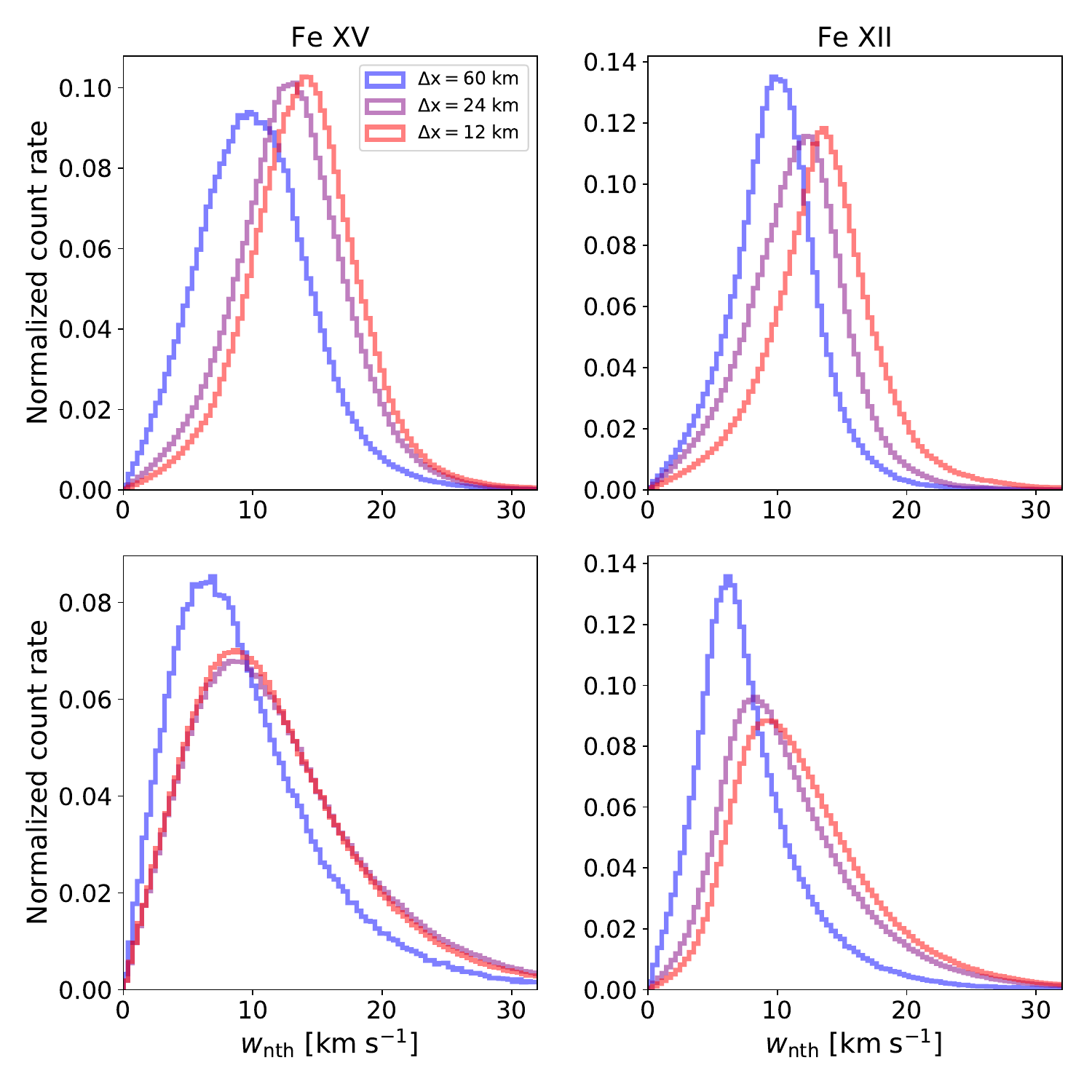}}
  \caption{Distribution of non-thermal velocity for numerical models with different resolutions for \fexv\ and \fexii\. Top row: Time-averaged normalized histograms for the non-thermal line width in \fexv\ and \fexii\ seen perpendicular to the guide field. We show the histograms for three different grid resolutions of 60 $\rm{km}$, 24 $\rm{km}$ and 12 $\rm{km}$, respectively. Bottom row: Time-averaged normalized histograms for the non-thermal line width seen parallel to the guide field.}
  \label{fig:hist_gridres}
\end{figure}

We find higher values of the non-thermal broadening at the peak of the distribution function and mean values for the medium and high resolution run (see Fig. \ref{fig:hist_gridres}).
For both emission lines, the difference between the 24 km run and the 12 km run is considerably smaller than the difference between the 60 km run and the 24 km run, suggesting that 12 km is an adequate resolution to study non-thermal line widths. This holds true for both the perpendicular and the parallel LOS. Even for the HR run, however, the line width does not seem to have completely converged. For higher numerical resolutions, we therefore expect slightly higher line broadening, which would also bring the results closer to observations.
For the \fexv\ emission, the peak of the distribution lies at 6.7 $\rm{km\; s^{-1}}$ for the lowest resolution of 60 km, 13.1 $\rm{km\; s^{-1}}$ for the medium resolution and 14.1 $\rm{km\; s^{-1}}$ for the highest resolution run. The mean values for the distribution are 10.3 $\rm{km\; s^{-1}}$, 13.3 $\rm{km\; s^{-1}}$ and 13.9 $\rm{km\; s^{-1}}$, respectively.\\
For the emission in the \fexii\ line, the peaks of the histograms are at slightly higher values of 11 $\rm{km\; s^{-1}}$, 13.4 $\rm{km\; s^{-1}}$, and 14.8 $\rm{km\; s^{-1}}$. 
We find average non-thermal line widths of 11.1 $\rm{km\; s^{-1}}$, 14 $\rm{km\; s^{-1}}$, and 15.2 $\rm{km\; s^{-1}}$ for the three different resolutions.
Seen parallel to the magnetic guide field, the average non-thermal line broadening at the footpoints in the \fexv\ line is 9.4, 11.8 and 11.5 $\rm{km\; s^{-1}}$ and the peaks are at 6.7, 8.1 and 8.5 $\rm{km\; s^{-1}}$ in the \fexv\ emission.
The histogram peak values for the \fexii\ emission are at 6, 8.1 and 9.2 $\rm{km\; s^{-1}}$ and the average values are 7.7, 10.8 and 11.9 $\rm{km\; s^{-1}}$, respectively.\\

\section{Velocity power spectra}

\begin{figure}
\resizebox{\hsize}{!}{\includegraphics{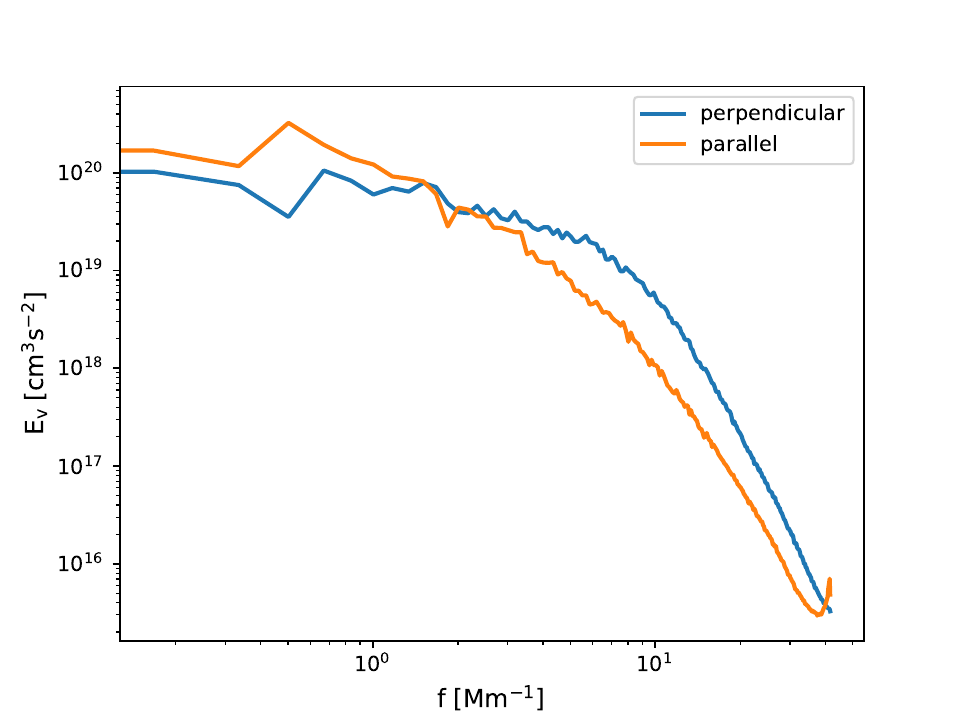}}
  \caption{Power spectra of the velocity components perpendicular and parallel to the magnetic guide field averaged over a slab with 1 Mm thickness centered on the loop apex for run HR at 22.21 min heating. For a discussion see Sect. \ref{section:res_dep}.}
  \label{fig:fourier}
\end{figure}

The velocity power spectra associated with transverse and axial velocities shown in Fig. \ref{fig:fourier} are calculated as
\begin{align*}
    E_{\rm{perp}}(k_{x,y})&=\frac{1}{\Delta k_{x,y}}\int_{k_{x,y}\leq \sqrt{k_{x}^2+k_{y}^2}<k_{x,y}+\Delta k_{x,y}}|\hat{v}_{\rm{perp}}|^2 dk_{x} dk_{y},\\
    E_{\rm{axial}}(k_{x,y})&=\frac{1}{\Delta k_{x,y}}\int_{k_{x,y}\leq \sqrt{k_{x}^2+k_{y}^2}<k_{x,y}+\Delta k_{x,y}}|\hat{v}_{\rm{axial}}|^2 dk_{x} dk_{y},
\end{align*}
where $\hat{v}_{\rm{perp}}$ and $\hat{v}_{\rm{axial}}$ are the Fourier transforms of the velocity field perpendicular and parallel to the loop axis.

\end{appendix}

\bsp	
\label{lastpage}
\end{document}